\documentclass[sigplan,10pt]{acmart}
\usepackage{graphicx}
\usepackage{booktabs}
\usepackage{xspace}
\usepackage{amsmath}
\usepackage{url}
\usepackage{listings}
\usepackage{xcolor}
\usepackage{hyperref}
\usepackage{subcaption}
\usepackage[capitalize]{cleveref}
\usepackage{enumitem}
\usepackage{siunitx}

\usepackage{algorithmicx}
\usepackage{algpseudocode}
\usepackage{algorithm}

\usepackage{tabularx}
\usepackage{makecell}
\usepackage{ragged2e}

\usepackage{multirow}
\usepackage{balance}

\newcommand{\project}{TokenCake\xspace}

\newcommand{\KVC}{KV Cache\xspace}
\newcommand{\KVCs}{KV Caches\xspace}
\newcommand{\KVCc}{KV-Cache-centric\xspace}

\newcommand{\spatialscheduler}{Spatial Scheduler\xspace}
\newcommand{\temporalscheduler}{Temporal Scheduler\xspace}

\newcommand{\FC}{Function Call\xspace}
\newcommand{\FCs}{Function Calls\xspace}
\newcommand{\fc}{function call\xspace}
\newcommand{\fcs}{function calls\xspace}
\newcommand{\fcparadigm}{\textit{LLM Inference1 $\Rightarrow$ \FC $\Rightarrow$ LLM Inference2}\xspace}

\newcommand{\fcstart}{call\_start\xspace}
\newcommand{\fcfinish}{call\_finish\xspace}

\acmBadgeR[https://www.acm.org/publications/policies/artifact-review-and-badging-current]{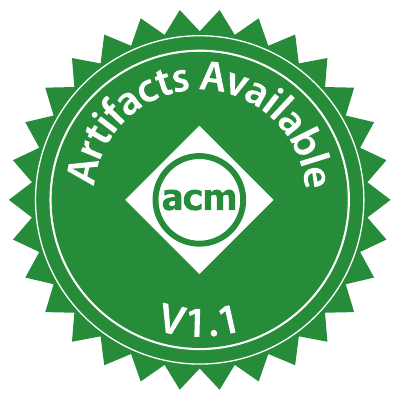}
\acmBadgeR[https://www.acm.org/publications/policies/artifact-review-and-badging-current]{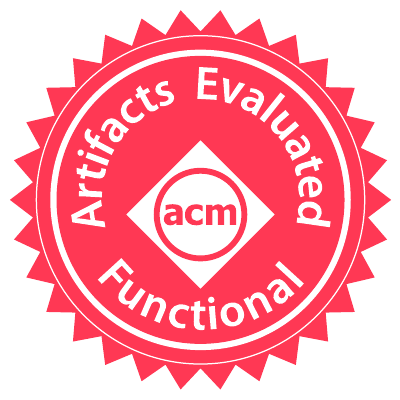}
\acmBadgeR[https://www.acm.org/publications/policies/artifact-review-and-badging-current]{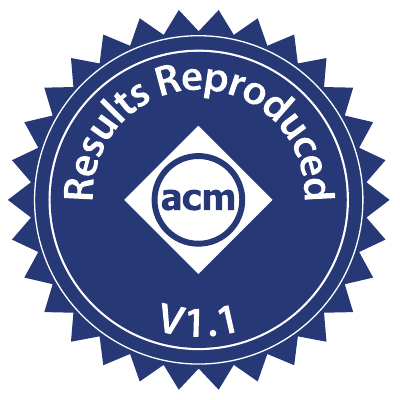}

\begin{document}

\copyrightyear{2027}
\acmYear{2027}
\setcopyright{cc}
\setcctype{by}
\acmConference[EuroSys '27]{22nd European Conference on Computer Systems}{April 19-23, 2027}{Rabat, Morocco}
\acmBooktitle{22nd European Conference on Computer Systems (EuroSys '27), April 19-23, 2027, Rabat, Morocco}
\acmDOI{10.1145/3842654.3848547}
\acmISBN{979-8-4007-2971-3/2027/04}

\title{\project: A KV-Cache-centric Serving Framework for LLM-based Multi-Agent Applications}

\author{Zhuohang Bian}
\orcid{0009-0007-3377-8926}
\affiliation{
  \institution{Peking University}
  \city{Beijing}
  \country{China}
}
\email{zhbian26@stu.pku.edu.cn}

\author{Feiyang Wu}
\orcid{0009-0002-6638-509X}
\affiliation{
  \institution{Peking University}
  \city{Beijing}
  \country{China}
}
\email{fywu25@stu.pku.edu.cn}

\author{Zhuoran Li}
\orcid{0009-0001-8049-1190}
\affiliation{
  \institution{Peking University}
  \city{Beijing}
  \country{China}
}
\email{zhuoranli@stu.pku.edu.cn}

\author{Teng Ma}
\orcid{0000-0002-7104-1526}
\affiliation{
  \institution{Alibaba Group}
  \city{Beijing}
  \country{China}
}
\email{stmatengss@gmail.com}

\author{Youwei Zhuo}
\orcid{0000-0002-1557-2613}
\authornote{Corresponding author.}
\affiliation{
  \institution{Peking University}
  \city{Beijing}
  \country{China}
}
\email{youwei@pku.edu.cn}



\begin{abstract}
Large Language Models (LLMs) are increasingly deployed in complex multi-agent applications that rely on external \fcs.
This workload creates severe performance challenges for the \KVC: spatial contention leads to the eviction of critical agents' caches and temporal underutilization leaves the cache of agents stalled on long-running \fcs idling in GPU memory.

We present \project, a \KVCc serving framework that bridges this gap by co-optimizing scheduling and memory management through an agent-aware design.
\project's \temporalscheduler employs an event-driven, opportunistic policy to proactively offload idle \KVCs during \fcs and uses predictive uploading to hide data transfer latency.
\project's \spatialscheduler uses dynamic memory partitioning, guided by a hybrid priority metric combining graph structure and runtime state, to reserve GPU memory for critical-path agents.
Our evaluation on representative multi-agent benchmarks shows that \project reduces end-to-end latency by up to 47.06\% in memory-constrained settings and improves effective GPU utilization by 16.9 percentage points compared to vLLM.
\project is publicly available at \url{https://github.com/pku-lemonade/TokenCake}.

\end{abstract}

\maketitle

\section{Introduction}
\label{sec:intro}

Large Language Models (LLMs) are powerful reasoning engines, and applications built upon them are evolving from single-response generation to complex, multi-agent systems.
This evolution has enabled powerful applications in domains like autonomous code generation~\cite{swebench,sweagent,openhands}, complex financial analysis~\cite{tradingagents}, and realistic environment simulation~\cite{agentsociety}.
The defining characteristic of these applications is a dual-interaction model: frequent external \textbf{agent-tool} interaction through \fcs, and complex internal \textbf{agent-agent} collaboration through structured workflows.
Externally, agents use \fcs to interact with tools, data sources, and APIs.
Internally, multiple specialized agents collaborate within a dependency graph to solve a larger problem.

\begin{figure}[!b]
\centering
\begin{minipage}[t]{\linewidth}
\centering
\begin{subfigure}[b]{0.48\linewidth}
\centering
    \includegraphics[width=\linewidth]{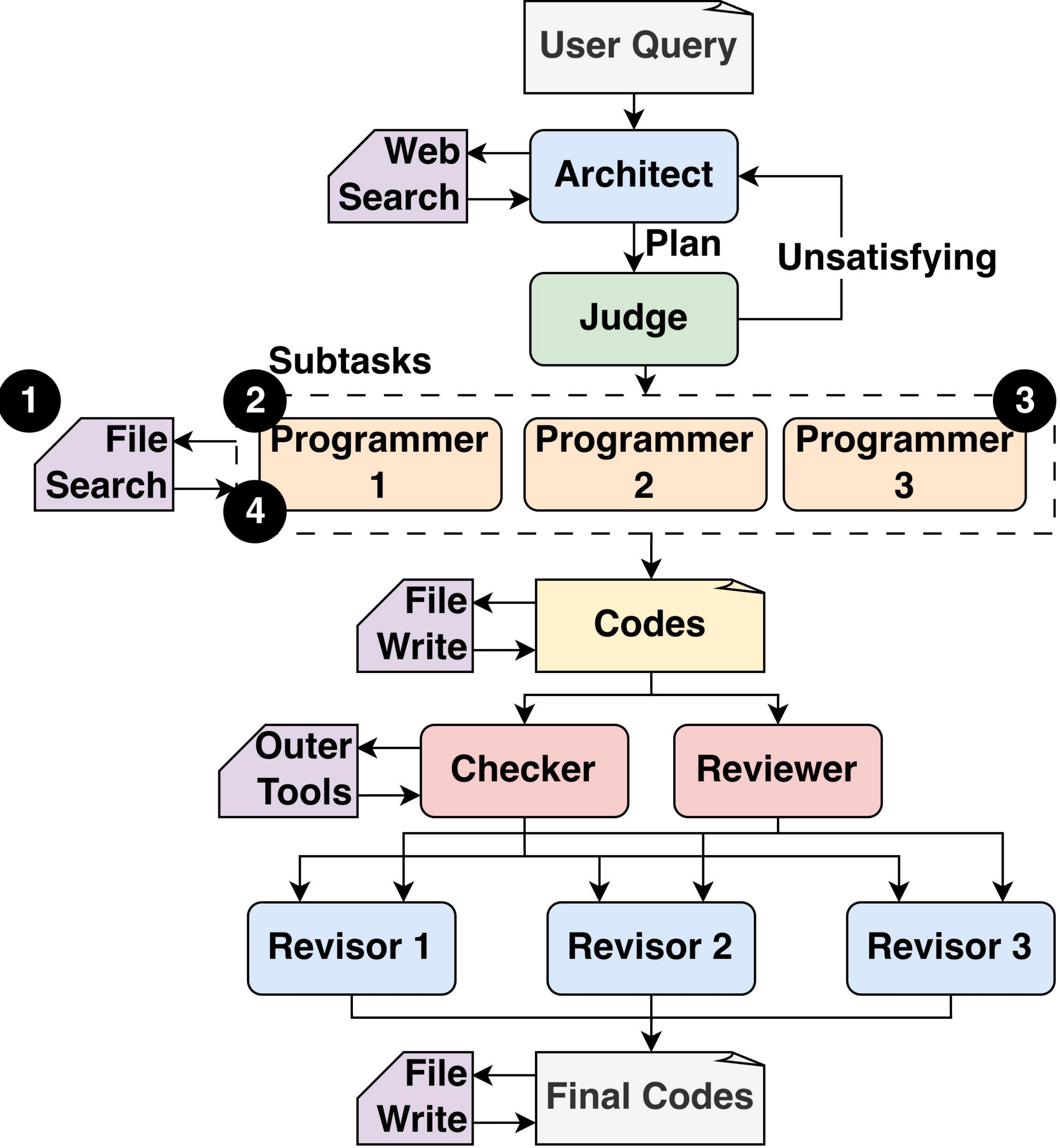}
    \caption{Code-Writer}
    \label{fig:multi_agent_coding}
\end{subfigure}
\begin{subfigure}[b]{0.48\linewidth}
    \centering
    \includegraphics[width=\linewidth]{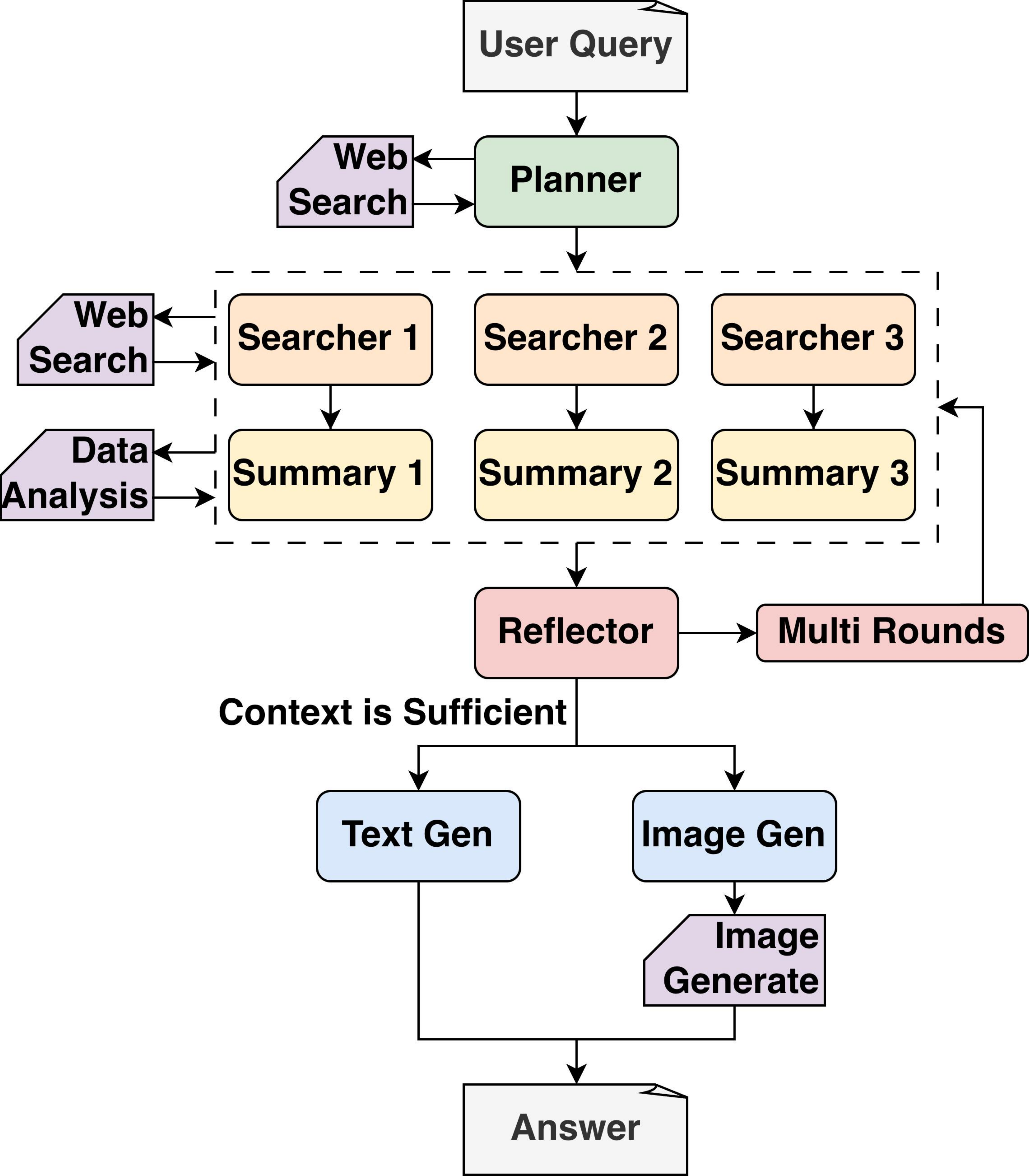}
    \caption{Deep-Research}
    \label{fig:multi_agent_research}
\end{subfigure}
\end{minipage}
\caption{
Example LLM-based Multi-Agent Applications.
Each colored box represents a specialized agent.
Purple boxes denote \fcs to external tools.
}
\Description{Two DAG diagrams of multi-agent applications. Left (Code-Writer): User Query flows to Architect (with Web Search), Judge, three parallel Programmers (with File Search), Codes (with File Write), Checker (with Outer Tools) and Reviewer in parallel, three Revisors, and Final Codes. Right (Deep-Research): User Query flows to Planner (with Web Search), three parallel Searcher-Summary pairs (with Web Search and Data Analysis), Reflector with multi-round loop, then Text Gen and Image Gen (with Image Generate) converging to Answer. Purple boxes denote function calls to external tools.}
\label{fig:agent_apps}
\vspace{-10pt}
\end{figure}

\begin{figure}[!t]
\centering
\begin{minipage}[t]{0.48\linewidth}
\centering
\begin{subfigure}[t]{\textwidth}
\includegraphics[width=\linewidth]{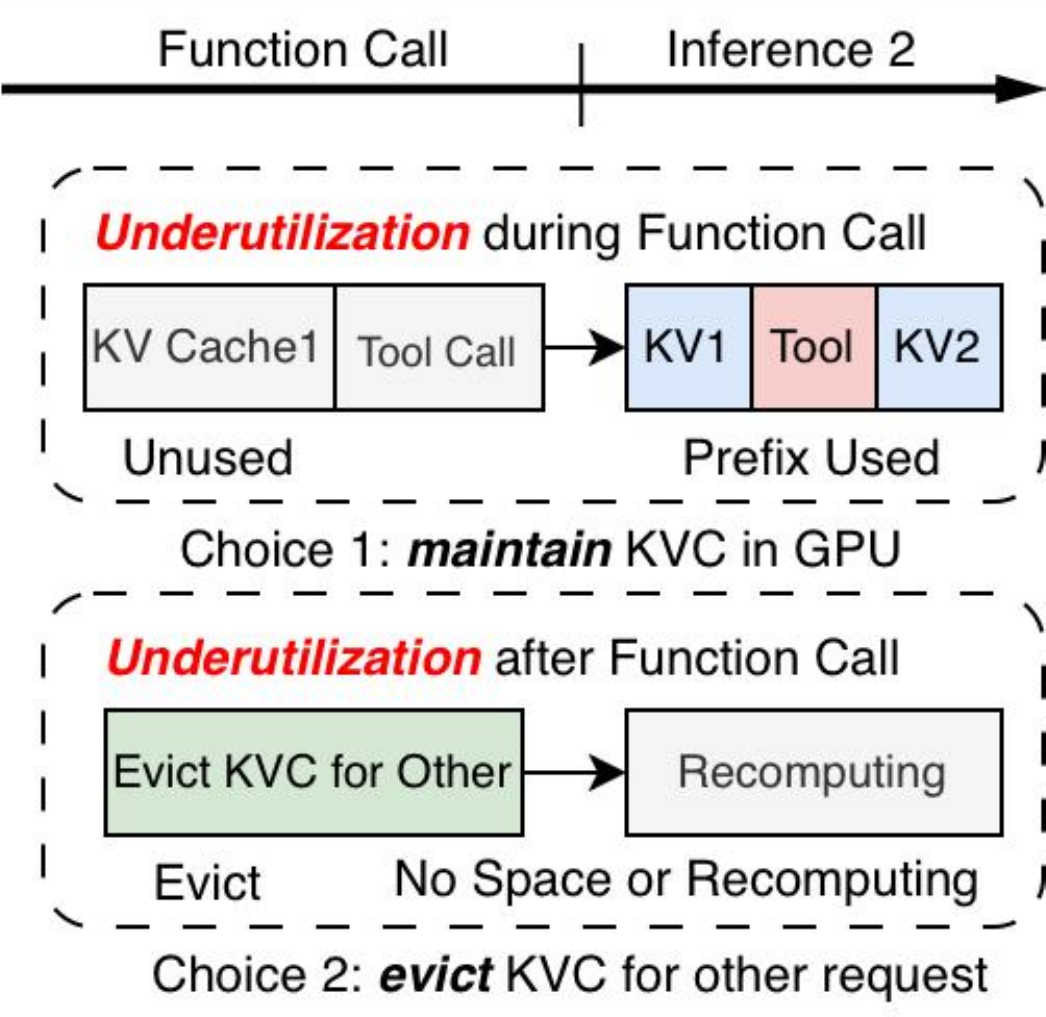}
\caption{
The lifecycle of an agent's \KVC during a \fc.
}
\label{fig:utilization_model}
\end{subfigure}
\end{minipage}
\hfill
\begin{minipage}[t]{0.48\linewidth}
\centering
\begin{subfigure}[t]{\textwidth}
\includegraphics[width=\linewidth]{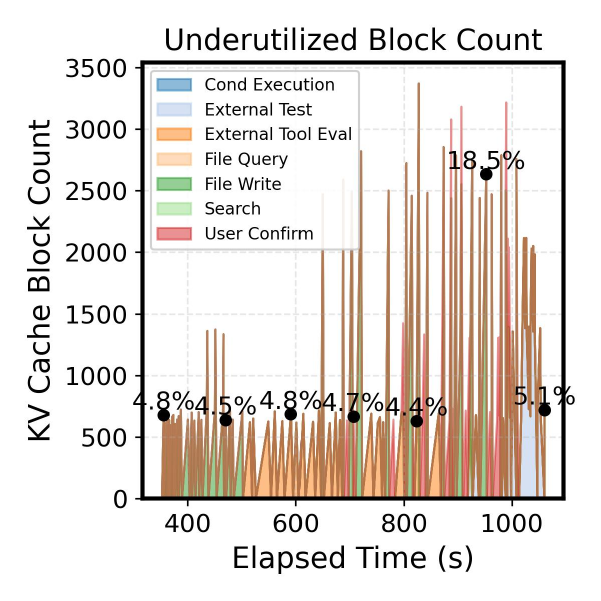}
\caption{
Idle \KVC blocks due to external \fcs.
}
\label{fig:utilization_exp}
\end{subfigure}
\end{minipage}
\caption{The Temporal Underutilization Problem.}
\Description{Left: schematic showing two choices during a function call -- Choice 1 maintains KV Cache in GPU (wasting memory), Choice 2 evicts KV Cache for other requests (requiring costly recomputation). Right: time-series plot of underutilized KV Cache block counts over elapsed time, with color-coded bars for different function call types (File Query, External Tool Eval, Search, etc.) and GPU utilization percentages (4.4\%--18.5\%) annotated at intervals, showing large numbers of blocks sitting idle during external function calls.}
\label{fig:insight_utilization}
\vspace{-5mm}
\end{figure}

Figure~\ref{fig:agent_apps} illustrates this application model with two representative examples.
Code-Writer orchestrates programmers, reviewers, and testers that inspect repositories, edit files, and repeatedly invoke shells and test harnesses~\cite{metagpt,chatdev,swebench,sweagent,openhands,anthropic_claude_code,openai_codex}.
Deep-Research interleaves reasoning with search, page navigation, source extraction, and report drafting over long execution horizons~\cite{google_gemini_deep_research,openai_deep_research,anthropic_claude_research,storm,costorm,webthinker,mind2web2,gaia}.
The combination of complex internal dependencies and frequent, long-running external interactions produces workload patterns that are fundamentally different from traditional LLM inference.
These patterns introduce critical performance challenges for the underlying serving infrastructure that existing systems are not designed to handle.

These challenges are rooted in inefficient \KVC management, which manifests along two dimensions: \textbf{temporal underutilization} and \textbf{spatial contention}.

Temporal underutilization roots in the frequent and long-running \fcs inherent to agentic workloads.
An agent's execution follows an \fcparadigm pattern, during which its \KVC---the exact prefix needed for the subsequent inference---sits idle in GPU memory.
Figure~\ref{fig:utilization_model} illustrates this lifecycle: after the first inference phase, the agent's \KVC remains resident while it waits for the external tool to respond.
This forces a difficult trade-off: retain the cache and waste GPU resources that could serve active requests, or evict it and incur a costly recomputation when the agent resumes.
This inefficiency is substantial.
As shown in Figure~\ref{fig:utilization_exp}, at peak moments, as much as $18.5$\% of the GPU \KVC pool is occupied by stalled agents, directly reducing the system's capacity for active computation.
Systems like Ayo~\cite{teola} have identified the latency challenge posed by this pattern and propose a workflow-level optimization.
Ayo intelligently pipelines the execution of LLM and non-LLM micro stages, aiming to overlap the tool-use-time of one agent with the computational work of another.
While this approach effectively hides latency by improving the application's end-to-end execution schedule, it is fundamentally compute-centric.
Ayo's scheduler optimizes the flow of operations but remains blind to the state of the underlying GPU memory resources.
Consequently, the \KVC of the stalled agent continues to idly occupy valuable GPU memory, a problem that is exacerbated when multiple agents stall concurrently.

\begin{figure}[!t]
\centering
\begin{minipage}[t]{0.48\linewidth}
\centering
\begin{subfigure}[t]{\linewidth}
\centering
\includegraphics[width=\linewidth]{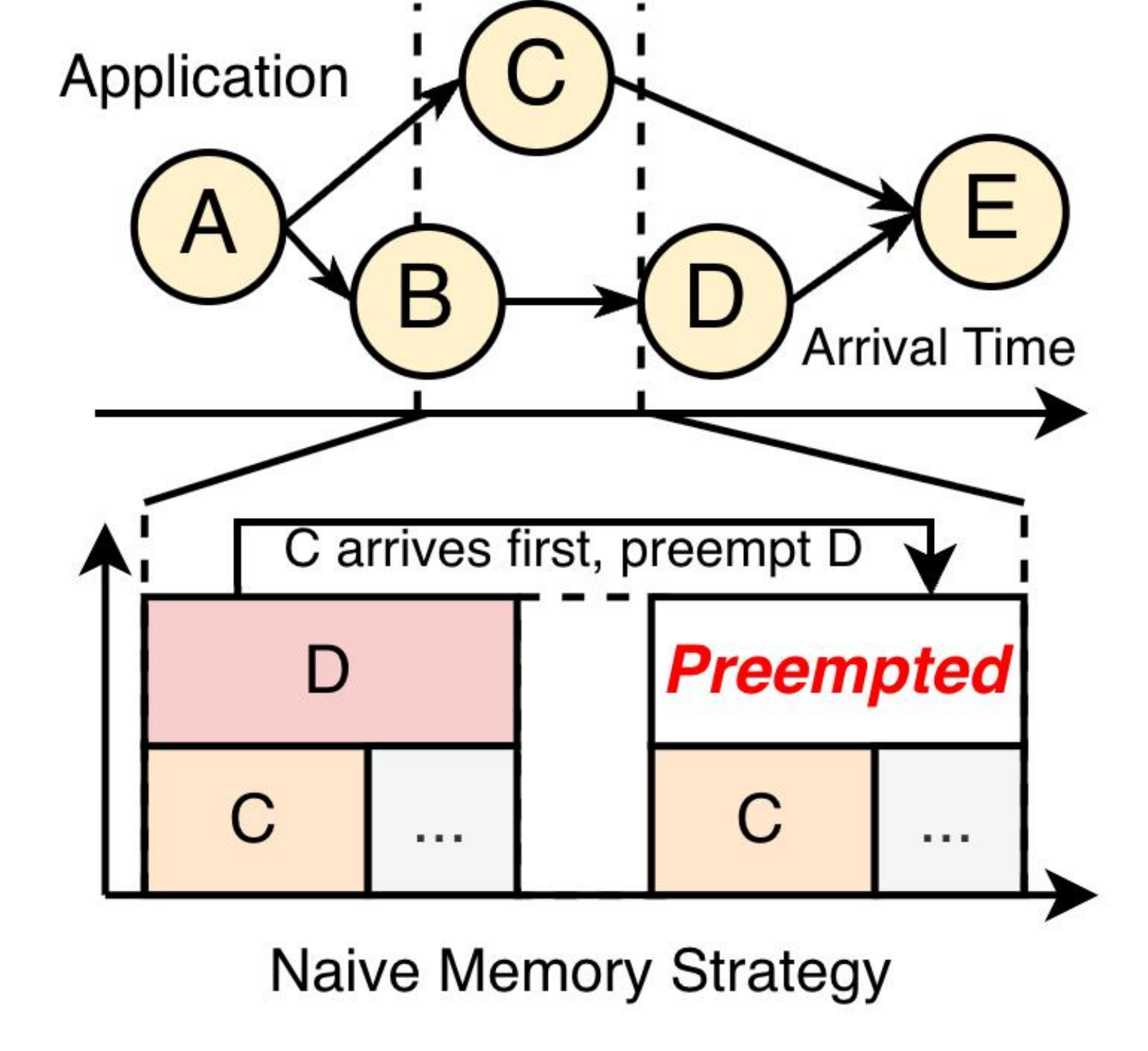}
\caption{
\KVC blocks held by non-critical agents.
}
\label{fig:contention_model}
\end{subfigure}
\end{minipage}
\hfill
\begin{minipage}[t]{0.48\linewidth}
\centering
\begin{subfigure}[t]{\linewidth}
\centering
\includegraphics[width=\linewidth]{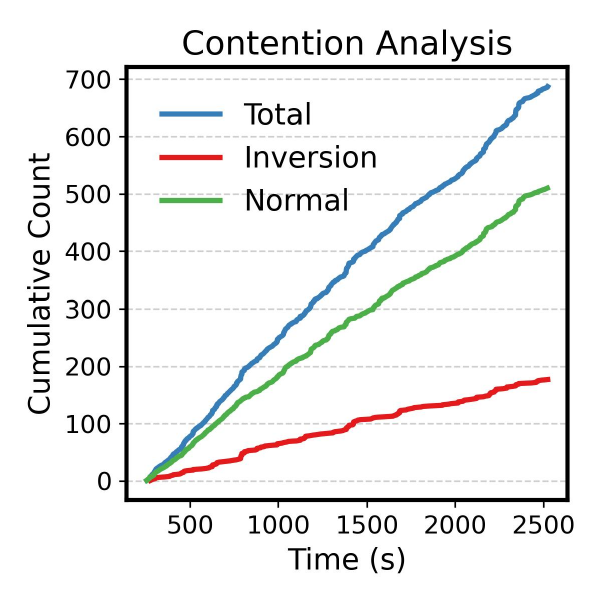}
\caption{
Preemption events over time of the Code-Writer workload.
}
\label{fig:contention_exp}
\end{subfigure}
\end{minipage}
\caption{The Spatial Contention Problem.}
\Description{Left: diagram showing an application DAG (A fans out to B and C, B leads to D, C and D converge on E) with a naive memory strategy timeline where agent C arrives first, occupies GPU memory, and preempts agent D (a critical-path agent). Right: line chart of cumulative preemption counts over time showing Total (reaching 690), Normal (510), and Inversion (180) events, demonstrating that priority inversions account for roughly 26\% of all preemptions in the Code-Writer workload.}
\label{fig:insight_contention}
\vspace{-7mm}
\end{figure}

Spatial contention arises as numerous agents compete for limited GPU memory.
As illustrated in Figure~\ref{fig:contention_model}, an agent on the critical path can be stalled if a non-critical agent arrives earlier and occupies the limited GPU memory.
Agent-unaware allocation policies like FCFS lead to a problem we term \textit{critical inversion}: a non-critical agent occupies memory and causes the eviction of a critical-path agent's \KVC.
This forces the evicted critical agent to undergo a costly context recomputation, stalling the entire application workflow.
We measure this phenomenon on the Code-Writer application and count the critical inversion events over time.
As shown in Figure~\ref{fig:contention_exp}, these harmful preemptions occur frequently under realistic workloads.
While systems like Parrot~\cite{parrot} and Agentix~\cite{agentix} are workflow-aware, their optimizations operate at the request level, agnostic to the fine-grained \KVC contention between individual agents.
They optimize the order and batching of requests but do not manage the underlying memory allocation.
Consequently, even with an optimal schedule, a high-throughput, non-critical task group identified by Parrot could still occupy GPU memory and inadvertently cause the eviction of a latency-sensitive, critical agent's \KVC.
This exposes a fundamental limitation: scheduling optimization alone cannot solve memory contention.

To address these challenges, we present \project, a \KVCc serving framework that co-optimizes scheduling and memory management through an agent-aware design.
\project begins with a frontend API that allows users to define an application's internal agent collaborations and external tool interactions as a graph (\S\ref{sec:frontend_api}).
This graph enables two specialized schedulers to manage the \KVC lifecycle with application-level context.
To mitigate temporal underutilization, the \textbf{\temporalscheduler} uses an event-driven, opportunistic policy to proactively offload the \KVC of stalled agents during \fcs and uses predictive uploading to hide transfer latency (\S\ref{sec:temporal-scheduler}).
To resolve spatial contention, the \textbf{\spatialscheduler} employs a dynamic memory partitioning policy, guided by a hybrid priority metric, to reserve memory for critical-path agents (\S\ref{sec:spatial-scheduler}).

The two schedulers share a unified pressure-aware coordination protocol that captures GPU capacity, reserved capacity, waiting demand, offloadable stalled blocks, and pending upload debt.
This shared protocol prevents the two schedulers from optimizing against different notions of pressure and ensures that every memory movement is justified by a concrete scheduling benefit.

Our evaluation on representative multi-agent benchmarks (\S\ref{sec:evaluation}) shows that \project reduces end-to-end latency by up to 47.06\% compared to vLLM in memory-constrained settings and improves GPU utilization by 16.9 percentage points under high load.
Ablation studies confirm that agent-aware scheduling and temporal offload must work together: agent scheduling alone reduces critical-path latency under memory pressure, while selective offloading helps when \fc stalls expose a useful scheduling window.
Neither mechanism subsumes the other, and applying offload without agent context can degrade performance.

\section{Background And Motivation}
\label{sec:background}

\subsection{LLM-Based Multi-Agent Applications}

\noindent\textbf{Agents.} A common design pattern in agentic systems is the decomposition of complex problems into a series of sub-tasks handled by specialized agents~\cite{autogen,metagpt,agentscope,sweagent,openhands,costorm}.
These collaborations are often modeled as a Directed Acyclic Graph (DAG), where nodes represent agents and edges represent explicit dependencies.
In these workflows, the dependencies are critically important; for instance, a reviewer agent cannot begin its task until a programmer agent has completed its code generation.
Furthermore, not all agents contribute equally to the application's end-to-end latency.
Some agents lie on the critical path, meaning any delay they experience directly increases the total time-to-result for the entire application.

\noindent\textbf{\FCs.}
A key feature of modern agents is the use of \fcs to interact with external data sources or execute actions~\cite{react,toolformer}.
These calls connect the LLM to a vast array of tools, such as database clients, code interpreters, or third-party web APIs~\cite{toolllm,bfcl,mcpmark,agentsysbench}.
To standardize these interactions, the Model Context Protocol (MCP)~\cite{mcp} is emerging as an open standard, providing a unified interface and a rich ecosystem of pre-defined tools.
As shown in Table~\ref{tab:mcp_tools_latency}, compiled from the official MCP documentation and our empirical measurements, the latencies of these common tools have a wide and often unpredictable distribution.

\begin{table}[t]
\centering
\small
\caption{Latency characteristics of common tools in MCP.}
\label{tab:mcp_tools_latency}
\begin{tabular}{@{}llll@{}}
\toprule
\textbf{Tool} & \textbf{Device} & \textbf{Latency} & \textbf{Variability} \\
\midrule
File System & CPU & 100\,ms & 50\,ms \\
Git & CPU & 100\,ms & 100\,ms--1\,s \\
Database (SQLite) & CPU & 100--1000\,ms & 500\,ms \\
Web Search & CPU & 1--5\,s & 1--10\,s \\
AI Generation & GPU & 5--30\,s & 10--60\,s \\
\bottomrule
\end{tabular}
\end{table}

\subsection{Limitations Of Existing Serving Systems}

The challenges of spatial contention and temporal underutilization expose the limitations of two distinct categories of state-of-the-art serving systems: those that are agent-aware but compute-centric, and those that are \KVCc but agent-agnostic.

A class of recent work has made serving systems agent-aware by incorporating the application's DAG into the scheduling logic.
Systems like Parrot~\cite{parrot} and Agentix~\cite{agentix} use this graph to mitigate head-of-line blocking by prioritizing critical requests, while Ayo~\cite{teola} optimizes the execution pipeline for an individual agent's interaction with external tools.

While these approaches improve high-level orchestration, their focus is fundamentally compute-centric.
They optimize the order and batching of requests but do not manage the underlying memory allocation.
Consequently, they cannot prevent critical inversion, as a high-throughput but non-critical task can still occupy GPU memory and cause the eviction of a critical agent's \KVC.
Furthermore, because they are not memory-centric, they do not address temporal underutilization, lacking the mechanisms to manage or repurpose the idle \KVC of stalled agents.

Another line of work has focused on making serving more \KVCc, introducing advanced memory management and offloading policies.
For instance, vLLM\cite{vllm}'s PagedAttention solves internal memory fragmentation, while systems like Mooncake\cite{mooncake} and CachedAttention\cite{cachedattention} have implemented offloading for general workloads.

However, while these systems are memory-aware, their policies are fundamentally agent-agnostic.
They treat all \KVC with equal importance, lacking the context to differentiate a critical-path agent from a non-critical one, which leaves them vulnerable to critical inversion.
Furthermore, their offloading policies are typically reactive---triggered by memory pressure or session inactivity---rather than proactive.
They are not designed to leverage the predictable idle periods during \fcs to mitigate underutilization.

\section{Overview}
\label{sec:system-overview}

\begin{figure}[t]
\centering
\includegraphics[width=\linewidth]{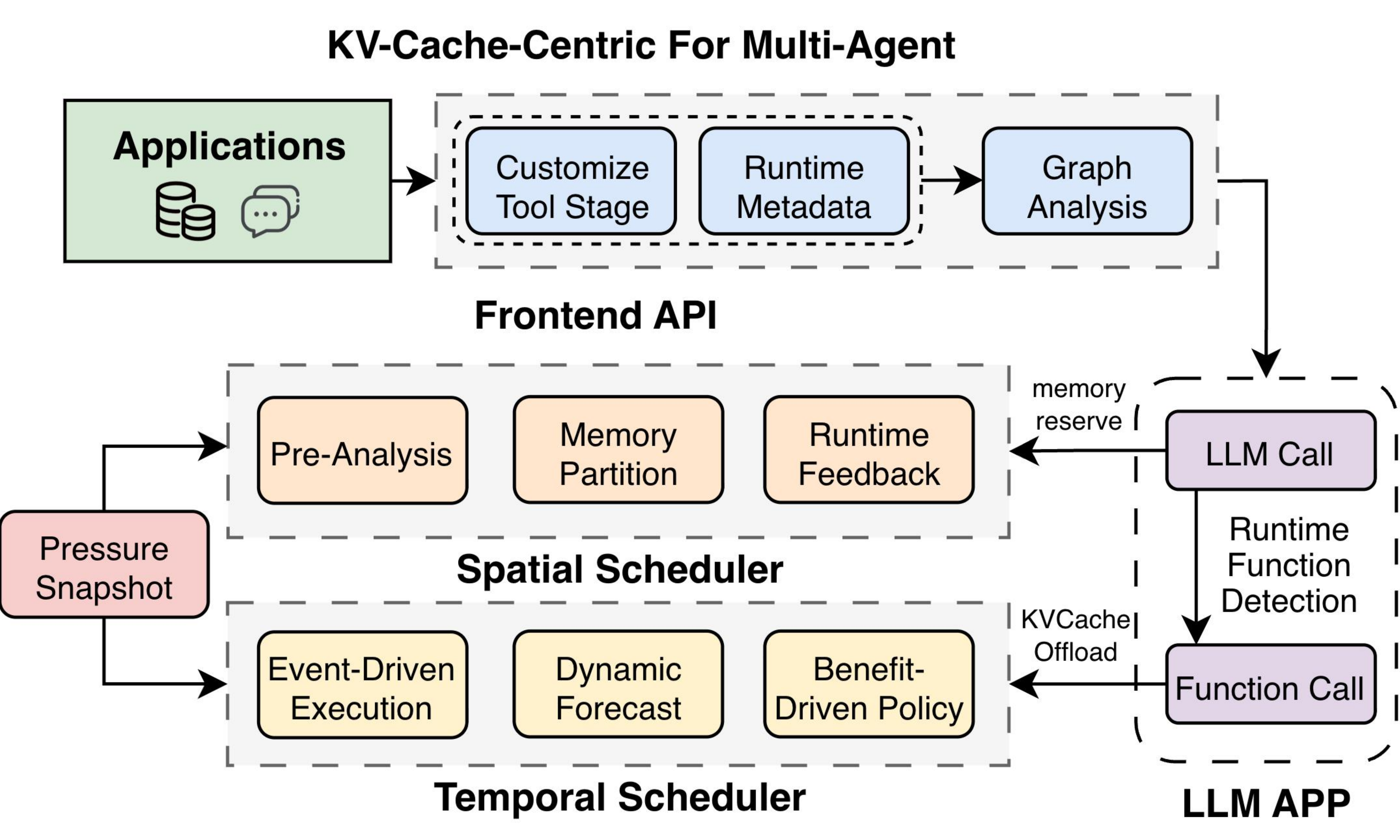}
\vspace{-5mm}
\caption{\project Overview.}
\Description{System architecture diagram. Applications connect to a Frontend API (Customize Tool Stage, Runtime Metadata, Graph Analysis). Below, the Spatial Scheduler (Pre-Analysis, Memory Partition, Runtime Feedback) and Temporal Scheduler (Event-Driven Execution, Dynamic Forecast, Benefit-Driven Policy) manage KV Cache resources. On the right, the LLM APP flow (LLM Call, Runtime Function Detection, Function Call) triggers memory reserve to the Spatial Scheduler and KVCache Offload to the Temporal Scheduler. A Pressure Snapshot feeds into the Spatial Scheduler.}
\label{fig:system_overview}
\vspace{-5mm}
\end{figure}

\project is designed to optimize multi-agent application performance by managing \KVC resources across both temporal and spatial dimensions.
Figure~\ref{fig:system_overview} shows the three primary components of \project: a Frontend API, a \temporalscheduler, and a \spatialscheduler.

The Frontend API (\S\ref{sec:frontend_api}) translates a user's multi-agent application into a graph annotated with \fc stages and time estimates.
The \temporalscheduler (\S\ref{sec:temporal-scheduler}) consumes this graph to manage the lifecycle of individual \KVC blocks over time, offloading idle caches during \fcs and uploading them before the agent resumes.
The \spatialscheduler (\S\ref{sec:spatial-scheduler}) partitions GPU memory dynamically based on both the static graph structure and runtime signals, reserving capacity for critical-path agents so that non-critical work cannot evict their \KVCs.
Both schedulers are coordinated through a shared memory pressure-aware coordination protocol that gives them a consistent view of GPU capacity and prevents contradictory memory decisions.

\subsection{Frontend API}
\label{sec:frontend_api}

\project provides a programming interface that lets users describe a multi-agent application as a Directed Acyclic Graph (DAG).
Nodes represent agents or computational units, and edges represent data dependencies.
The API exposes three kinds of information that existing serving systems lack: graph structure, fine-grained \fc stages, and performance metadata.

\begin{figure}[t]
\centering
\includegraphics[width=0.9\linewidth]{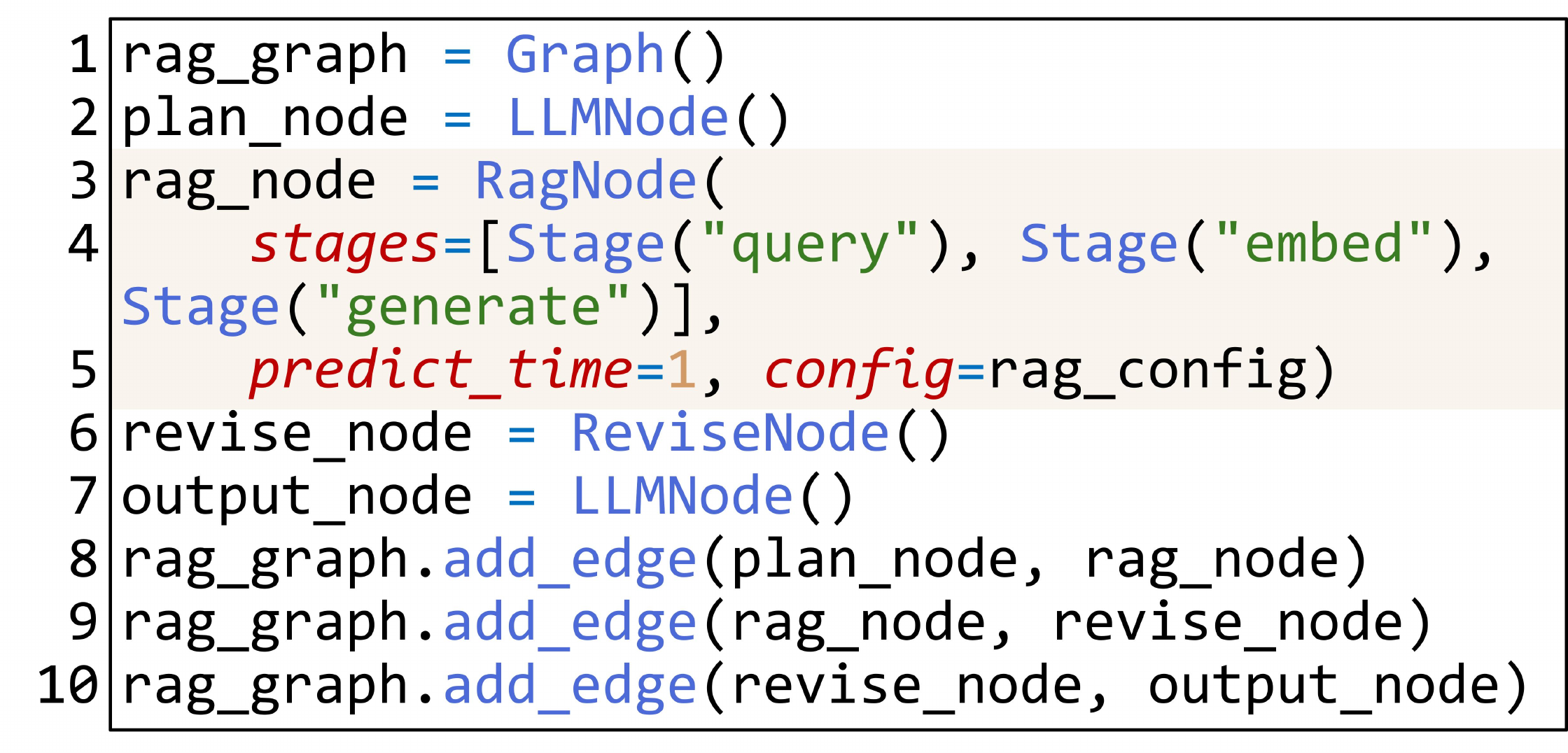}
\vspace{-5mm}
\caption{
Defining a multi-agent RAG application with the \project API.
}
\Description{A 10-line Python code listing. It creates a Graph, defines an LLMNode (plan), a RagNode with three stages (query, embed, generate) and predict\_time=1, a ReviseNode, and an output LLMNode, then connects them sequentially with add\_edge calls. The RagNode definition (lines 3--5) is highlighted, showing fine-grained stage decomposition and predicted execution time metadata.}
\label{fig:front_api}
\vspace{-5mm}
\end{figure}

Figure~\ref{fig:front_api} shows how a user builds a simple Retrieval Augmented Generation application.
The \texttt{FuncNode} abstraction (line~4) lets users decompose a \fc into multiple sequential stages.
Each stage emits its own \textit{\fcstart} and \textit{\fcfinish} events, so the \temporalscheduler sees the progress of a long \fc rather than a single start-to-finish interval and can time the upload more precisely.
Without decomposition, \project treats the \fc as one opaque stage, and only the upload timing becomes coarser.
The API also allows users to supply critical performance metadata directly within the graph definition.
In line~5, the \texttt{predict\_time} parameter provides an estimated execution time for the \fc, which helps the \temporalscheduler make more accurate decisions about when to offload and prefetch an agent's \KVC.

\subsection{Coordination between the Spatial and Temporal Schedulers}
\label{sec:scheduler_coordination}

\begin{figure}[t]
\centering
\includegraphics[width=\columnwidth]{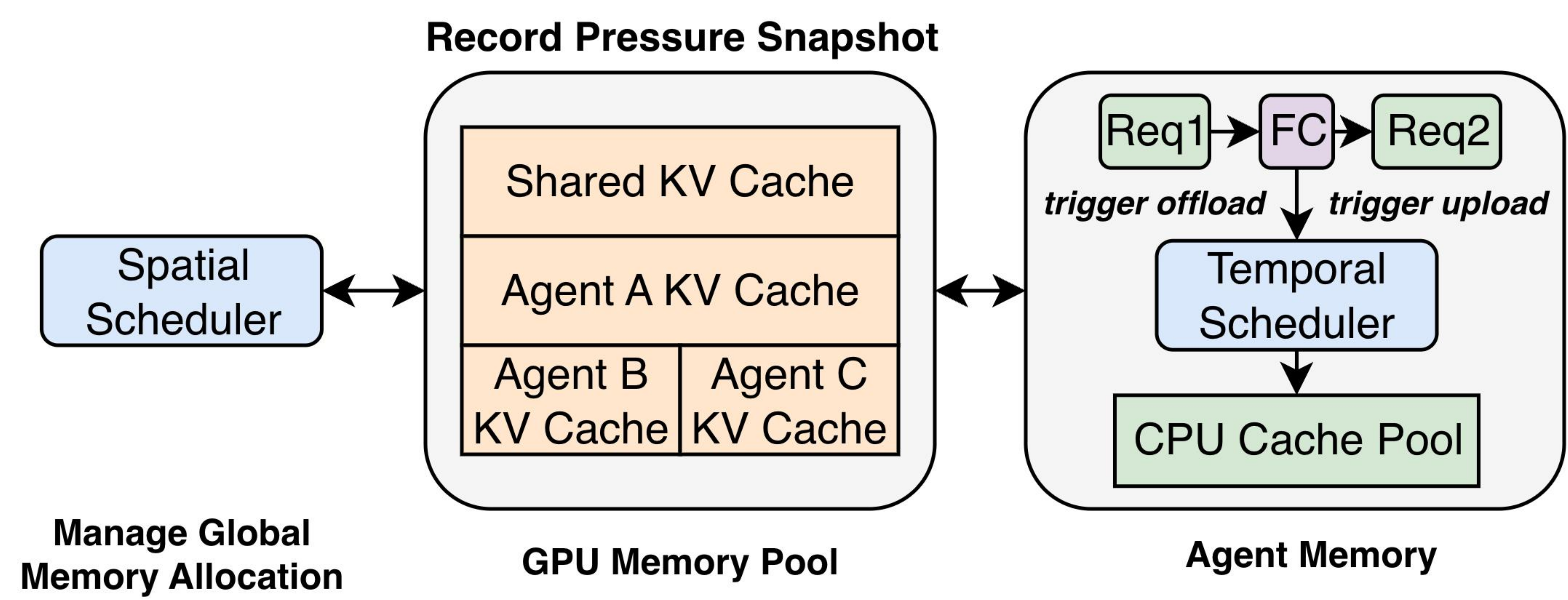}
\caption{
Coordination between the \spatialscheduler and the \temporalscheduler within a single scheduling step.
}
\Description{Diagram showing the GPU Memory Pool (center) divided into Shared KV Cache, Agent A KV Cache, and smaller blocks for Agents B and C. The Spatial Scheduler (left) manages global memory allocation. On the right, Agent Memory shows a request flow (Req1, FC, Req2) with the Temporal Scheduler receiving trigger offload and trigger upload signals, connecting to a CPU Cache Pool below. The top label reads Record Pressure Snapshot.}
\label{fig:scheduler_interaction}
\vspace{-5mm}
\end{figure}

The \temporalscheduler and the \spatialscheduler optimize different dimensions of \KVC management but compete for the same GPU memory.
Without coordination, the \temporalscheduler may upload a \KVC into blocks reserved for a critical request, or the \spatialscheduler may admit a request into blocks needed for an imminent upload.
\project prevents such conflicts through a shared pressure snapshot and a fixed execution order within each scheduling step, as illustrated in Figure~\ref{fig:scheduler_interaction}.

The pressure snapshot captures GPU and CPU block availability, per-agent-type reserved capacity, waiting demand, offloadable stalled blocks, and pending upload debt.
Both schedulers read this snapshot so that every memory movement has a concrete scheduling benefit: an offload occurs only when freed blocks can admit useful work, and an upload occurs only when the resumed request will not displace a more important active one.

Each scheduling step proceeds in four phases:
(1) refresh application metadata and build the pressure snapshot;
(2) update the \spatialscheduler's reservation plan if the adjustment window has expired;
(3) the \temporalscheduler reserves GPU blocks for imminent uploads, transfers ready CPU-resident \KVCs back to the GPU, and evaluates newly stalled requests for offload;
(4) the \spatialscheduler forms the next batch under agent-aware admission control, routing each waiting request to shared capacity, reserved capacity, or deferral.

\noindent\textbf{Running Example.}
Consider the Code-Writer application (Figure~\ref{fig:multi_agent_coding}): \textcircled{\small 1}~Programmer~1 invokes a \texttt{FileSearch} tool call while Programmer~3 waits in the queue for GPU \KVC blocks.
\textcircled{\small 2}~The \temporalscheduler detects that Programmer~1's predicted stall exceeds the round-trip transfer cost and that a waiting request can use the freed blocks, so it offloads Programmer~1's \KVC to CPU memory.
\textcircled{\small 3}~Concurrently, the \spatialscheduler identifies Programmer~3 as high-priority (high structural importance and synchronization pressure from a pending join) and admits it into the freed blocks, while deferring a lower-priority background task to prevent critical inversion.
\textcircled{\small 4}~As the predicted completion time approaches, the \temporalscheduler uploads Programmer~1's \KVC back to the GPU, ensuring it resumes without delay.
The search result is appended to Programmer~1's own context, so the restored \KVC remains a valid prefix and needs no recomputation.
Neither decision makes sense alone: temporal offload without spatial awareness frees blocks indiscriminately, while spatial reservation without temporal offload lacks the freed capacity to protect.

\section{The \temporalscheduler}
\label{sec:temporal-scheduler}

When an agent issues a \fc, its \KVC sits idle in GPU memory until the external tool responds.
The \temporalscheduler converts this idle period into a productive scheduling window by offloading the stalled agent's \KVC to CPU memory and uploading it back before the agent resumes, avoiding the costly recomputation that a naive eviction would require.
This reuse is valid because the tool result is appended to the context of the agent that issued the call, so the saved \KVC remains a prefix when the same agent resumes.
\project therefore reuses a \KVC only within one agent; downstream agents in the DAG build their own \KVCs beyond any prefix-cache hit.

Table~\ref{tab:kvc_offload_comparison} contrasts \project's policy with three prior systems.
CachedAttention~\cite{cachedattention} offloads on session inactivity;
Mooncake~\cite{mooncake} offloads under memory pressure or LRU eviction;
InferCept~\cite{infercept} reactively swaps \KVCs to CPU upon external call interception;
Pensieve~\cite{pensieve}, HCache~\cite{hcache}, NEO~\cite{neo}, and multi-turn dialogue serving~\cite{multiturndialogue} reduce restoration or offload costs through session persistence or I/O--computation overlap.
None use the \fc event as an explicit scheduling trigger, none differentiate caches by graph criticality, and none gate offload on whether freed blocks can admit useful work.
\project avoids unnecessary overhead through its opportunistic policy (\S\ref{sec:opportunistic-policy}), which offloads only when freed memory will be productively used.

\begin{table*}[ht]
\centering
\small
\setlength{\tabcolsep}{3pt}
\caption{
Comparison of \KVC offloading and prefetching policies.
}
\label{tab:kvc_offload_comparison}
\begin{tabular}{l l c c c c c}
\toprule
\textbf{Category} & \textbf{Criteria} & \textbf{\project{}} & \textbf{Mooncake} & \textbf{CachedAttention} & \textbf{InferCept} & \textbf{Others\textsuperscript{$\dagger$}} \\
\midrule
\multirow{2}{*}{\textbf{General}}
  & FC Aware & \textbf{Yes} & No & No & Partial & No \\
  & Data Granularity & Block & Block & Layer & Block & Varies \\
\midrule
\multirow{3}{*}{\textbf{Offload}}
  & Strategy & \textbf{Proactive} & Reactive & Reactive & Reactive & Reactive \\
  & Trigger & FC Start & Cache Pool Pressure & Session Inactive & Interception Signal & Memory Pressure \\
  & Decision Logic & Cost-Benefit & LRU & Session & Min-Waste Heuristic & Heuristic \\
\midrule
\multirow{3}{*}{\textbf{Prefetch}}
  & Strategy & \textbf{Predictive} & Proactive & Reactive & Reactive & Reactive \\
  & Trigger & Predicted FC Completion & SLO-based Schedule & Session Resumption & Interception End & Session/I/O \\
  & Decision Logic & Static + Dynamic & Static & Static & FCFS & Static \\
\bottomrule
\multicolumn{7}{l}{\textsuperscript{$\dagger$}\small Pensieve~\cite{pensieve}, HCache~\cite{hcache}, NEO~\cite{neo}, Multi-turn Dialogue~\cite{multiturndialogue}.}
\end{tabular}
\end{table*}

\subsection{Event-Driven Offload And Predictive Upload}
\label{sec:event-driven-offload}

\begin{figure}[t]
\centering
\includegraphics[width=\columnwidth]{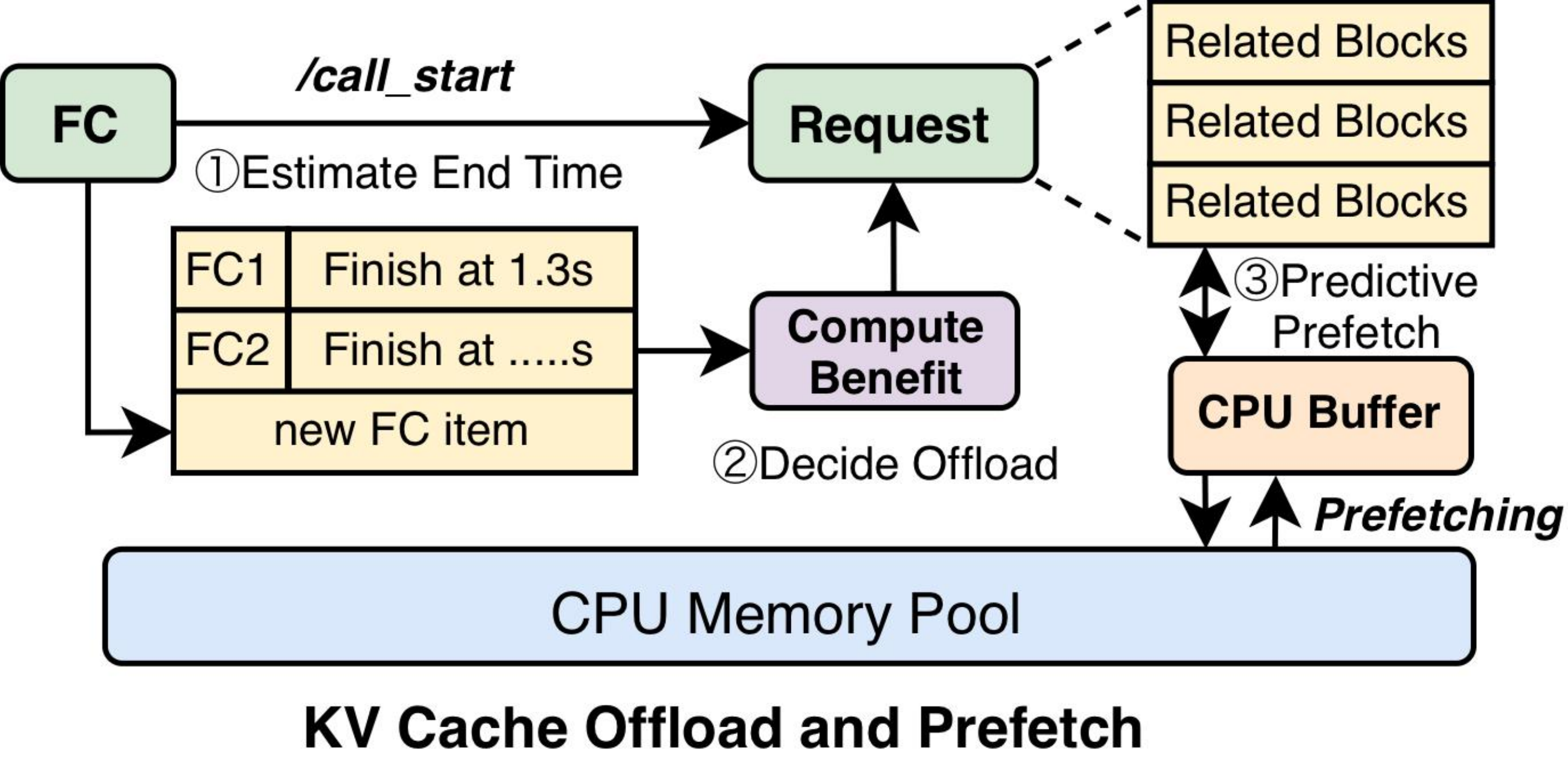}
\vspace{-5mm}
\caption{
Lifecycle of the \temporalscheduler's offload and predictive upload mechanism.
}
\Description{Flowchart showing three phases. Phase 1: a function call sends a /call\_start event and estimates end time. Phase 2: a Compute Benefit module decides whether to offload the KV Cache to a CPU Buffer. Phase 3: Predictive Prefetch retrieves related blocks from the CPU Buffer back to GPU ahead of the function call's completion. A CPU Memory Pool at the bottom stores offloaded KV cache blocks.}
\label{fig:temporal_pattern}
\vspace{-5mm}
\end{figure}

Figure~\ref{fig:temporal_pattern} illustrates the lifecycle that the \temporalscheduler manages for each stalled agent.
The scheduler is driven by two runtime events and a dynamic forecasting model that connects them.

\noindent\textbf{Runtime Event Loop.}
\label{sec:temporal_runtime_loop}
A \textit{\fcstart} event fires when an agent emits a \fc.
The scheduler consults the opportunistic policy (\S\ref{sec:opportunistic-policy}) to decide whether offloading the agent's \KVC is worthwhile.
If the policy approves, the \KVC is transferred to CPU memory asynchronously.
As the predicted completion time approaches, the scheduler begins a gradual upload (\S\ref{sec:priority-aware-upload}) to hide transfer latency behind other computation.
A \textit{\fcfinish} event fires when the tool result returns.
If the \KVC has already been uploaded, the agent resumes immediately.
If the tool returns earlier than predicted, the scheduler triggers an immediate upload to ensure correctness.
The observed execution time is fed back to the forecasting model to refine future predictions.

\noindent\textbf{Dynamic Forecasting.}
The scheduler maintains a historical estimate $t_{history}$ for each function type.
When no history is available, the estimate defaults to the value supplied by the user through the graph definition, or to a conservative system-wide constant if no user estimate exists.
Each observed execution time $t_{obs}$ updates the history as an exponentially weighted moving average, $t_{history} \leftarrow \alpha \cdot t_{obs} + (1-\alpha) \cdot t_{history}$, where $\alpha$ is the weight of the newest sample.
When the user also provides a time estimate ($t_{user}$), the model blends the two sources with a separate coefficient $\beta$:
\vspace{-2pt}
\begin{equation}
\label{eq:fc_time_estimate}
t_{estimate} = \beta \cdot t_{user} + (1 - \beta) \cdot t_{history}
\end{equation}

\subsection{Opportunistic Policy For Proactive Offloading}
\label{sec:opportunistic-policy}

Not every \fc stall justifies an offload.
Offloading is beneficial only when three conditions hold simultaneously: the predicted stall is long enough to cover a round-trip transfer, a waiting request exists that can use the freed GPU blocks, and the later upload can be prepared without displacing more important work.
The opportunistic policy encodes these conditions as a gate that rejects offloads when the net benefit is unclear.

\noindent\textbf{Transfer Overhead Estimation.}
The primary cost is the round-trip data transfer time between GPU and CPU memory.
For a request holding $N_{blocks}$ blocks, the transfer time is linear in $N_{blocks}$ in \project's scene:
\vspace{-2pt}
\begin{equation}
\label{eq:overhead_transfer}
T_{transfer} = T_{offload}(N_{blocks}) + T_{upload}(N_{blocks})
\end{equation}

Both terms are calibrated from measured per-block transfer constants on the target platform (\S\ref{sec:evo_offload_tradeoff}).

\noindent\textbf{Scheduling Window.}
The scheduler computes the available scheduling window as the predicted stall minus the transfer overhead:
$T_{window} = T_{FC} - T_{transfer}$.
If this window is non-positive, the offload is rejected immediately because the stall is too short.

\noindent\textbf{Waiting-Request Fit.}
As shown in Algorithm~\ref{alg:benefit_driven_offload}, the scheduler converts the scheduling window into a token capacity estimate using the system's observed decode throughput.
It then searches the waiting queue for a request whose incremental \KVC demand fits within the freed blocks and whose total work fits within the token capacity.
If no such request exists, offloading provides no immediate benefit and is rejected.
The search uses a \textit{first\_fit} policy that selects the first eligible request in queue order.
\S\ref{sec:evo_sensitivity} compares \textit{first\_fit} with \textit{best\_fit} and \textit{priority\_first} alternatives and shows that \textit{first\_fit} achieves the best balance between decision overhead and scheduling benefit.

\begin{algorithm}[htbp]
\small
\caption{Core Decision Logic of the \temporalscheduler}
\label{alg:benefit_driven_offload}
\begin{algorithmic}[1]
\Procedure{ShouldOffload}{req}
    \State $T_{transfer} \gets \text{EstimateTransferTime}(req.\mathit{blocks})$
    \State $T_{fc} \gets \text{PredictFCDuration}(\text{req.FC})$
    \If{$T_{fc} \le T_{transfer}$}
        \State \Return \textbf{false} \Comment{Stall too short.}
    \EndIf
    \State $T_{window} \gets T_{fc} - T_{transfer}$
    \State $N_{capacity} \gets T_{window} \times v_{throughput}$
    \State $waiting\_req \gets \text{FindFirstFitRequest}(N_{capacity})$
    \State \Return {$waiting\_req$ is not null}
\EndProcedure
\end{algorithmic}
\end{algorithm}
\vspace{-5mm}

\noindent\textbf{Hard Rejection And Soft Scoring.}
Before computing any score, the policy applies four hard rejections: CPU capacity is insufficient, the predicted stall is shorter than the transfer time, no waiting request fits within the freed blocks, or GPU memory pressure is below a configurable threshold.
If none of these conditions trigger, the policy computes a composite score that integrates GPU pressure, block fit quality, upload safety margin, and CPU capacity as positive signals, and penalizes offloading critical-path agents (using the \spatialscheduler's priority metric), near-completion requests, and requests with high migration churn.
The offload proceeds only if this score exceeds a threshold.
The dominant positive term rewards stalls that are long relative to transfer time; the dominant penalty discourages offloading agents that the \spatialscheduler has designated as critical.
An emergency exception allows offloading even high-importance requests under severe GPU pressure when the stall margin is large.

This multi-factor gate encodes the empirical finding from \S\ref{sec:component_controlled_studies}: CPU offload is not universally beneficial---it helps only when a stalled cache can be converted into useful active work within a safe scheduling window.

\subsection{Predictive Upload}
\label{sec:priority-aware-upload}

Predictive upload completes the temporal scheduling cycle.
If the upload starts only after a tool returns, the resumed request stalls on data transfer.
If the upload reserves all destination GPU blocks too early, active requests lose memory before the cache is needed.
\project resolves this tension through gradual reservation guided by both urgency and application importance.

The scheduler ranks upload candidates by $P_{upload} = I + U$, where $I$ is the normalized request importance from the \spatialscheduler's priority metric; $U$ is urgency based on proximity to the predicted \fc completion time.
Higher-priority and earlier-deadline requests are uploaded first.
At each scheduling step, the scheduler computes an upload budget that protects critical waiting requests:

\vspace{-2pt}
\begin{equation}
\label{eq:upload_budget}
B_{upload} = \max(0,\; B_{gpu}^{free} - \max(0,\; D_{critical} - B_{shared}^{free}))
\end{equation}

\noindent where $D_{critical}$ is the total block demand from critical waiting requests and $B_{shared}^{free}$ is the number of free blocks in the shared pool.
This ensures uploads never consume blocks that the \spatialscheduler needs for critical-path agents.

Rather than allocating all destination GPU blocks at once---which risks evicting active requests under high memory utilization---the scheduler reserves at most half of each candidate's remaining deficit per scheduling step:
\vspace{-2pt}
\begin{equation}
\label{eq:gradual_upload_reserve}
B_{reserve} = \min(B_{remain},\; \lceil \tfrac{B_{deficit}}{2} \rceil,\; B_{upload})
\end{equation}

\noindent This gradual approach amortizes allocation over several cycles, ensuring destination blocks are ready when the upload fires without displacing critical waiting work.
The upload priority uses the same importance signal as the \spatialscheduler's admission control (\S\ref{sec:spatial-scheduler}), so temporal and spatial decisions share a consistent notion of criticality.

\section{The \spatialscheduler}
\label{sec:spatial-scheduler}

The spatial contention problem arises because agents of different criticality compete for the same GPU \KVC memory.
A simple FCFS memory allocation policy allows a non-critical agent that arrives first to occupy blocks that a critical-path agent needs, forcing the critical agent into costly recomputation and stalling the entire application.
We call this phenomenon \textit{critical inversion}.
Prior agent-aware systems~\cite{parrot,agentix,hermes} recognize the importance of priority scheduling for LLM applications, but they operate at the request scheduling level and do not manage \KVC memory allocation.
Consequently, even with an optimal schedule, a non-critical agent can still exhaust GPU blocks and evict a critical agent's \KVC.
Preble~\cite{preble} and KVCache Cache in the Wild~\cite{kvcachewild} coordinate reuse with load or history, but neither reserves GPU blocks according to application criticality.

\begin{figure}[t]
\centering
\includegraphics[width=\columnwidth]{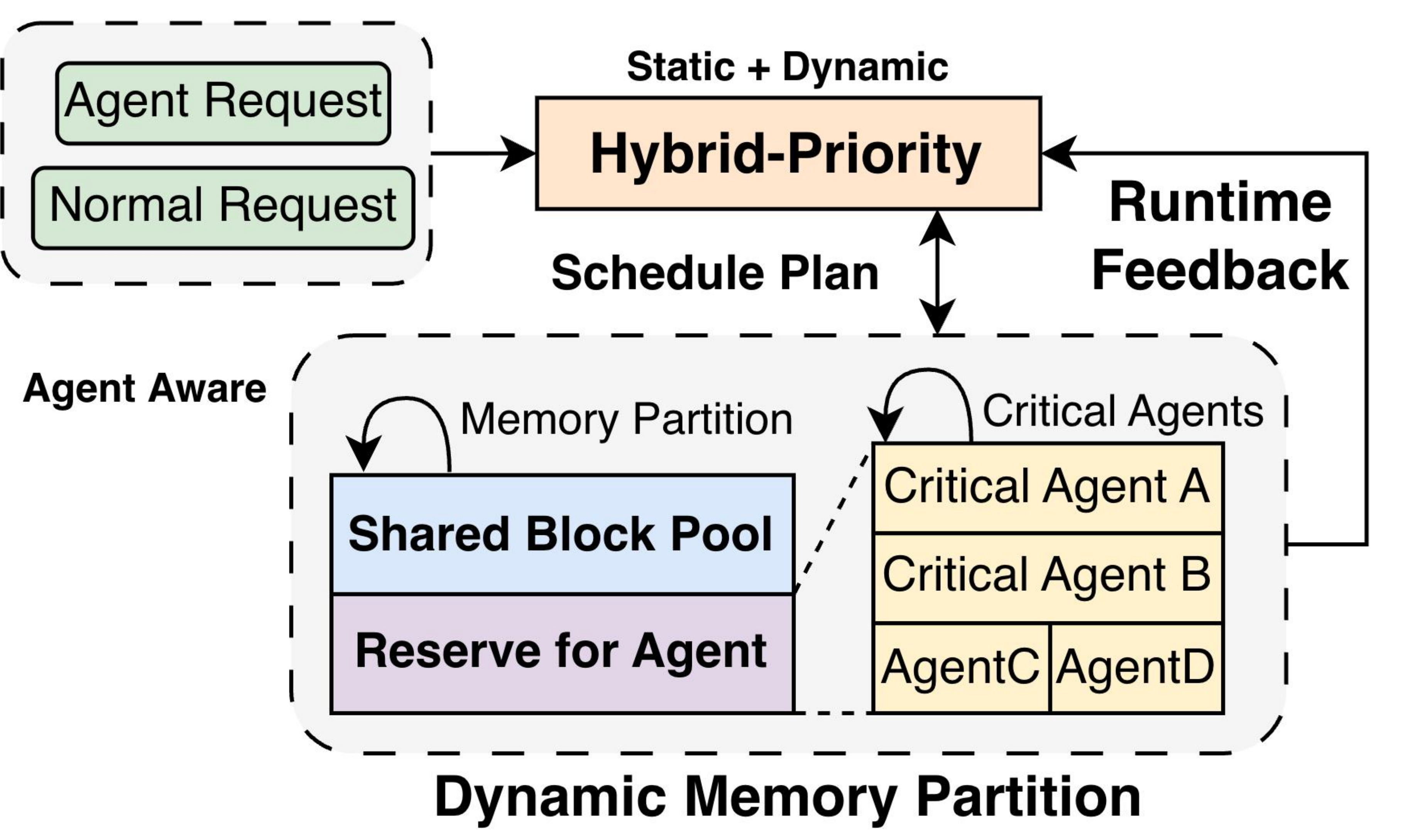}
\caption{
The \spatialscheduler's dynamic memory partitioning feedback loop.
}
\Description{Diagram showing Agent Requests and Normal Requests feeding into a Hybrid-Priority module (combining static and dynamic signals) that produces a Schedule Plan. Below, a Dynamic Memory Partition area contains a Shared Block Pool and Reserve for Agent pool on the left, plus a Critical Agents list (Agents A, B, C, D) on the right. A Runtime Feedback loop connects the schedule plan output back to the priority module.}
\label{fig:spatial_scheduler_working_pattern}
\vspace{-5mm}
\end{figure}

The \spatialscheduler solves critical inversion at the memory level rather than only at the scheduling level.
It controls which agents can allocate GPU \KVC blocks by dynamically partitioning memory into a shared pool that is available to all agents and a reserved pool that is accessible only to designated critical agents.
This partitioning protects capacity for critical-path agents under memory pressure, ensuring that non-critical work cannot evict their \KVCs.
The partition sizes adapt to runtime conditions through a feedback loop that combines static graph signals with dynamic runtime signals, as illustrated in Figure~\ref{fig:spatial_scheduler_working_pattern}.
The \spatialscheduler periodically re-evaluates which agent types are critical, adjusts the total reserved capacity based on GPU memory pressure, and distributes reserved blocks among critical types proportionally to their importance and memory footprint.

\noindent\textbf{Multi-GPU Support.}
\label{sec:multi-gpu-support}
\project supports the same reservation policy in multi-GPU deployments.
The \spatialscheduler maintains per-GPU shared and reserved pools and uses the same agent priority metric to coordinate admission across devices.
For tensor-parallel inference, a request is admitted only when the required \KVC blocks can be reserved on all participating GPUs.
The multi-GPU path keeps the policy unchanged and extends only the pressure snapshot with per-device free blocks, reserved blocks, and pending upload demand.

\subsection{Dynamic Memory Partitioning}
\label{sec:runtime_control}

The \spatialscheduler divides the GPU \KVC memory into two regions: a globally shared pool available to all agents, and a reserved pool accessible only to the most critical agents.
This two-pool structure protects critical work under contention while allowing non-critical work to proceed when memory is abundant.
As shown in Algorithm~\ref{alg:memory_reservation_update}, the partitioning adapts through a three-step update that runs periodically: adjusting the total reserved pool, selecting which agent types are critical, and distributing reserved capacity among them.

\noindent\textbf{Step 1: Adjusting The Total Reserved Pool.}
\label{sec:adjust_reserved_pool}
The scheduler monitors GPU block usage and adjusts the fraction of total blocks allocated to the reserved pool.
When usage is high, the reserved fraction increases to protect critical agents from contention.
When usage is low, the reserved fraction decreases to return capacity to the shared pool.
This feedback prevents over-reservation during low-load periods and under-reservation during memory pressure.

\noindent\textbf{Step 2: Selecting Critical Agent Types.}
\label{sec:critical_agent_selection}
Before distributing reserved capacity, the scheduler determines which agent types are currently critical.
It computes the agent-type score $S_a$ (defined in \S\ref{sec:hybrid_priority_metric}) for every active agent type and designates the top fraction as critical.
The current implementation uses a critical-agent ratio of $0.75$ and caps the reserved pool at $30\%$ of total blocks.

\noindent\textbf{Step 3: Distributing Reserved Capacity Among Critical Agents.}
\label{sec:distribute_reserved_capacity}
Once the total reserved pool size and the critical agent set are determined, the scheduler partitions reserved blocks among individual critical agent types.
Each agent type's share is a weighted combination of two factors: its current GPU block usage (reflecting its actual memory footprint) and its priority score $S_a$ relative to other critical agents.
This ensures that agents which are both structurally important and memory-intensive receive proportionally larger reservations, while structurally important but memory-light agents still receive a non-zero protected allocation.

\begin{algorithm}[t]
\caption{Dynamic Memory Reservation Update}
\label{alg:memory_reservation_update}
\small
\begin{algorithmic}[1]
\Procedure{UpdateReservations}{}
    \State $N \gets$ TotalGpuBlocks()
    \State usage $\gets$ CurrentGpuUsage() / $N$

    \State \Comment{Step 1: Adjust reserved pool size}
    \If{usage $\ge$ high\_watermark}
        \State $\rho \gets \rho +$ step
    \ElsIf{usage $\le$ low\_watermark}
        \State $\rho \gets \rho -$ step
    \EndIf
    \State $\rho \gets$ Clamp($\rho$, $\rho_{\min}$, $\rho_{\max}$)

    \State \Comment{Step 2: Select critical agent types via $S_a$ (Eq.~\ref{eq:agent_type_score})}
    \State $\mathcal{C} \gets$ TopFraction(active\_types, $S_a$, critical\_ratio)

    \State \Comment{Step 3: Distribute reserved blocks}
    \ForAll{$a \in \mathcal{C}$}
        \State share $\gets$ $\frac{1}{2}\bigl($GpuUsage($a$)/$N$ $+$ $S_a / \textstyle\sum_{a' \in \mathcal{C}} S_{a'}\bigr)$
        \State reserved[$a$] $\gets$ share $\times$ $\rho \times N$
    \EndFor
\EndProcedure
\end{algorithmic}
\end{algorithm}

The current implementation uses a total reserved ratio that starts at $0.05$, increases by $0.05$ when GPU usage exceeds $0.75$, decreases by $0.05$ when usage falls below $0.40$, and is clamped to $[0.05, 0.30]$.
These thresholds make reservation conservative at low load and progressively stronger under memory pressure.

\subsection{Hybrid Priority Metric}
\label{sec:hybrid_priority_metric}

The \spatialscheduler uses two levels of priority.
A per-request priority $P_{req}$ orders individual requests in the scheduling queue and determines which requests enter the next batch.
A per-agent-type score $S_a$ determines which agent classes receive reserved \KVC capacity.
Both combine static graph signals with dynamic runtime signals, but they operate at different granularities and use different formulas.
Both metrics are enabled by the application-level context---the DAG structure, agent positions, and runtime history---that agent-agnostic systems lack.

\noindent\textbf{Per-Request Priority.}
The scheduler refreshes $P_{req}$ before every batch decision as a weighted sum over three dimensions:
\vspace{-2pt}
\begin{equation}
\label{eq:per_request_priority}
P_{req} = \alpha_{\text{struct}} \cdot f_{\text{struct}} + \alpha_{\text{sync}} \cdot f_{\text{sync}} + \alpha_{\text{aging}} \cdot f_{\text{aging}}.
\end{equation}

\emph{Structural importance} ($f_{\text{struct}}$) captures how much downstream work a request unlocks by combining the node's depth and in/out-degree into a single score.
\emph{Synchronization pressure} ($f_{\text{sync}}$) boosts straggler branches in parallel workflows.
At a join point, a lagging branch's priority increases inversely with its relative progress, preventing the merge node from becoming a bottleneck.
\emph{Temporal aging} ($f_{\text{aging}}$) prevents starvation and reduces tail latency by combining three signals: the fraction of the graph remaining, queue wait time, and a completion-pressure term that gives near-finished applications a final push.

\noindent\textbf{Agent-Type Score For Reservation.}
\label{sec:agent-score}
Unlike $P_{req}$, which scores individual requests, the per-agent-type score $S_a$ aggregates across all active requests of a given type to decide whether the class as a whole deserves memory reservation.
The score is a weighted sum over four dimensions:
\vspace{-2pt}
\begin{equation}
\label{eq:agent_type_score}
S_a = \max(1,\; w_1 P_a + w_2 U_a + w_3 H_a + w_4 G_a).
\end{equation}

\emph{Structural priority} ($P_a$) is a static priority, ensuring that a single high-criticality instance triggers protection for the entire type.
\emph{Runtime urgency} ($U_a$) captures how much the system has failed to serve type $a$, measured by preemption and waiting counts.
Preemption receives a larger coefficient because it directly signals \KVC capacity loss---the problem the \spatialscheduler targets.
\emph{Recomputation cost} ($H_a$) protects types whose \KVCs are expensive to rebuild by log-compressing average token count, execution time, and throughput.
\emph{Graph context} ($G_a$) captures the average structural position of type $a$'s active requests---depth and fan-in/out---ensuring that reservation reflects the agent type's role in the workflow DAG, not just its urgency or cost.

\section{Implementation}
\label{sec:implementation}

\project is a \KVCc serving system for multi-agent applications, composed of a front-end and an execution engine.
It is implemented in approximately 9k lines of Python code and reuses some components from vLLM.

\subsection{Frontend API}

The frontend extends vLLM's OpenAI Chat Completion endpoint with a stateful graph registration interface.
Users register a DAG of agents and \fc nodes before submitting inference requests.

Table~\ref{tab:fc_defines} lists the pre-built \texttt{FuncNode} types that the API provides.
Each type bundles a default execution-time estimate and internal stage decomposition.

\begin{table}[tbp]
\centering
\small
\caption{Pre-built \texttt{FuncNode} types in the frontend API.}
\label{tab:fc_defines}
\begin{tabular}{@{}ll@{}}
\toprule
\textbf{Node} & \textbf{Description} \\
\midrule
\texttt{FileReadNode} & Read the contents of a specified file. \\
\texttt{FileWriteNode} & Write content to a specified file. \\
\texttt{SearchNode} & Perform a web search query. \\
\texttt{FileQueryNode} & Query files under a specified path. \\
\texttt{DataAnalysisNode} & Multi-stage analysis of large datasets. \\
\texttt{UserConfirmNode} & Request user confirmation. \\
\texttt{ExternalTestNode} & Use external test tools. \\
\bottomrule
\end{tabular}
\vspace{-3mm}
\end{table}

\subsection{Function-Call Start And Finish Endpoints}

The execution engine exposes two HTTP endpoints that drive the \temporalscheduler.
When an application begins a \fc, it sends a request to \texttt{\fcstart} carrying the request identifier and an initial time estimate.
This event transitions the request into a stalled state and makes it eligible for offload evaluation.
When the \fc completes, the application sends a request to \texttt{\fcfinish} carrying the request identifier and the actual elapsed time.
This event marks the request as ready for upload and feeds the observed duration back to the per-function-type forecasting model in Equation~\ref{eq:fc_time_estimate}.

Both endpoints are processed asynchronously by a unified \texttt{MCPManager} that maintains per-request lifecycle state.
The manager maps each request to one of five states: running, pending-offload, offloaded, pending-upload, and uploaded.
The two endpoints and the scheduling loop described next drive all state transitions.

\subsection{CPU Migration Infrastructure}
\label{sec:impl_cpu_migration}

All \KVC migration is issued asynchronously on a dedicated stream.
Source GPU blocks are marked as pending-free immediately after the copy is issued and return to the free pool only after the transfer completes, preventing re-allocation of blocks still being read.

vLLM V1 removed host-memory swap support and relies on recomputation for evicted \KVCs.
\project re-introduces a CPU block pool that maintains a lightweight free list recycling fixed-size blocks without returning them to the operating system.
This avoids the costly system-allocator cycles that high-frequency offloading would otherwise induce, reducing worst-case CPU allocation latency from nearly a second to consistent sub-millisecond levels.

The block pool owns the mapping between GPU blocks, CPU blocks, and block hashes.
On offload, it links the GPU and CPU blocks, marks the \KVC as CPU-resident, and inserts the block hash into a CPU prefix-cache index.
On upload, it restores the GPU mapping and reuses the original GPU block when possible.
The CPU prefix-cache index extends the standard lookup path: a CPU hit avoids recomputation but creates an H2D transfer entry that must complete before the request can run.

\section{Evaluation}
\label{sec:evaluation}

This section answers five questions about \project.
First, does \project reduce end-to-end latency and improve GPU memory utilization compared to existing serving systems (\S\ref{sec:evo_headline})?
Second, how much does each component contribute, and must agent scheduling and temporal offload be coordinated (\S\ref{sec:component_controlled_studies})?
Third, how does \project compare with a remote-\KVC baseline and an agent-aware execution baseline (\S\ref{sec:evo_remote_agent_baselines})?
Fourth, how sensitive is the \temporalscheduler to tool-time variability and policy parameters (\S\ref{sec:evo_sensitivity})?
Fifth, is proactive offload practical and what overhead does it introduce (\S\ref{sec:evo_offload_tradeoff})?

\begin{figure*}[t!]
\centering
\includegraphics[width=\linewidth]{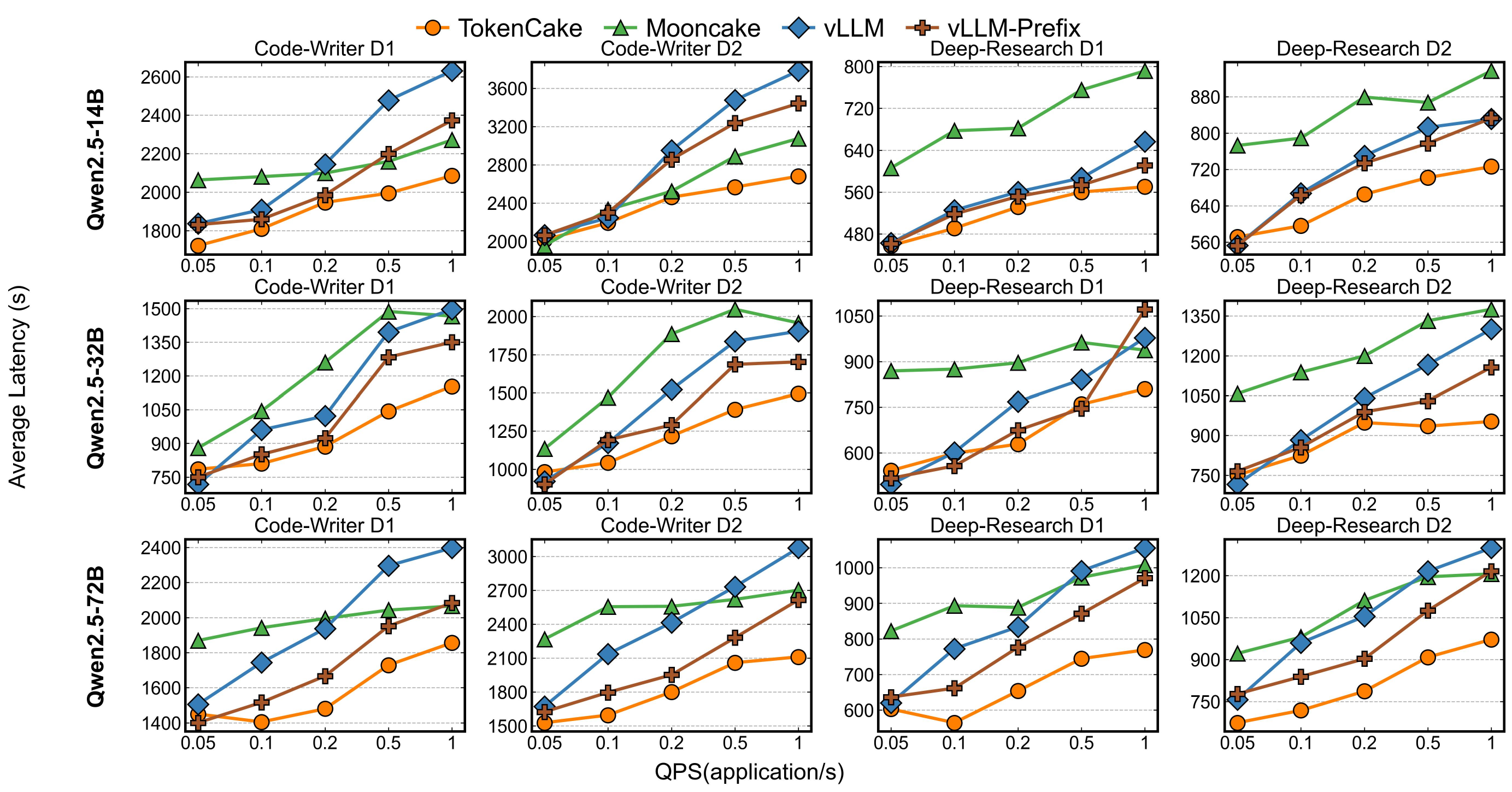}
\vspace{-5mm}
\caption{
End-to-end application latency comparison of \project, vLLM, vLLM-Prefix, and Mooncake.
Each chart plots average latency against queries-per-second (QPS) for the specified application, model, and dataset.
}
\Description{A 3x4 grid of line charts comparing average latency (seconds) vs. QPS (0.05--1.0 applications/s) for TokenCake, Mooncake, vLLM, and vLLM-Prefix. Rows: Qwen2.5-14B, 32B, 72B. Columns: Code-Writer D1, Code-Writer D2, Deep-Research D1, Deep-Research D2. TokenCake (orange circles) consistently achieves the lowest latency across nearly all configurations, with the gap widening at higher QPS levels.}
\label{fig:evo_qps_compare}
\end{figure*}

\subsection{Experimental Setup}
\label{sec:evo_setup}

\noindent\textbf{Hardware and Models.}
We evaluate \project on three configurations: Qwen2.5-14B on one NVIDIA A100 (80\,GB HBM), Qwen2.5-32B on one NVIDIA H20 (96\,GB HBM), and Qwen2.5-72B on two NVIDIA H20 GPUs with tensor parallelism degree 2.
The 72B configuration exercises \project's multi-GPU support, where the \spatialscheduler coordinates \KVC reservations across both GPUs.
All configurations reserve 100\,GB of CPU memory as the offload destination.

\noindent\textbf{Benchmark Applications.}
We implement two representative multi-agent applications (Figure~\ref{fig:agent_apps}).
Code-Writer (Figure~\ref{fig:multi_agent_coding}) orchestrates 11 agent types with frequent \fcs to file I/O, search, and external test tools, creating high memory pressure from many concurrent \KVC states.
Deep Research (Figure~\ref{fig:multi_agent_research}) models a research workflow with fewer agents but deeper dependency chains that stress critical-path optimization~\cite{gemini-deep-research}.

\noindent\textbf{Workload Generation.}
User requests are synthesized from the ShareGPT~\cite{sharegpt} and AgentCode~\cite{agentcode} datasets.
Application arrivals follow a Poisson distribution at varying rates.
External \fcs use a two-phase methodology.
We first execute each application with live tool endpoints and record both the tool outputs and latency distributions.
During evaluation, the tool server replays the recorded outputs while sampling durations from the collected distributions with a fixed seed, ensuring that all baselines receive identical tool behaviors and stall times for a fair comparison.
Latency ranges are listed in Table~\ref{tab:mcp_tools_latency}.

\noindent\textbf{Baselines.}
We compare against four systems.
\textbf{vLLM} (v0.8.6) is the standard \KVCc but agent-agnostic baseline.
\textbf{vLLM-Prefix} adds prefix caching to vLLM for shared prompt reuse.
\textbf{Mooncake} (v0.3.0-beta) is a remote-\KVC design that tests whether general-purpose disaggregated caching suffices without \fc awareness.
\textbf{Parrot} is an agent-aware, compute-centric system that tests whether scheduling alone solves memory contention.

\noindent\textbf{Metrics.}
We report end-to-end latency (average and tail percentiles) from application submission to final response, and GPU \KVC utilization as the fraction of \KVC blocks occupied over time.

\noindent\textbf{Parameter Configuration.}
Table~\ref{tab:default_params} lists the default parameters used in all experiments.
We use a single fixed configuration across all workloads without per-application tuning.
Graph terms drive request priority, while bounded aging prevents starvation; the application start-offset term is capped at 60 seconds.
Similarity is the inverse count of same-type nodes at the same depth, and parallelism is that count minus one.

\begin{table}[t]
\centering
\small
\caption{Default parameter configuration.}
\label{tab:default_params}
\vspace{-3mm}
\begin{tabular}{@{}lll@{}}
\toprule
\textbf{Component} & \textbf{Parameter} & \textbf{Default} \\
\midrule
\multirow{5}{*}{\makecell[l]{\spatialscheduler}}
  & Critical-agent ratio & 0.75 \\
  & Reserved ratio range & [0.05, 0.30] \\
  & Reserve adjustment step & $\pm$0.05 \\
  & GPU usage low watermark & 0.40 \\
  & GPU usage high watermark & 0.75 \\
\midrule
\multirow{5}{*}{\makecell[l]{\temporalscheduler}}
  & Request selection policy & first\_fit \\
  & Min GPU usage for offload & 0.60 \\
  & Benefit score threshold & 1.0 \\
  & History EWMA weight $\alpha$ & 0.5 \\
  & User--history blend $\beta$ (Eq.~\ref{eq:fc_time_estimate}) & 0.5 \\
\midrule
\multirow{5}{*}{\makecell[l]{Request score $P_{req}$}}
  & Importance / depth & 100 / 15 \\
  & Out- / in-degree & 45 / 10 \\
  & Similarity / remaining depth & 20 / 20 \\
  & Parallelism / start offset & 120 / 6 \\
  & Age / wait / completion & 5 / 8 / 5 \\
\multirow{2}{*}{\makecell[l]{Type score $S_a$}}
  & Weights $w_1{:}w_2{:}w_3{:}w_4$ & 2:1:0.5:1 \\
  & $U_a$: preemption coefficient & 1.5 \\
\bottomrule
\end{tabular}
\end{table}

\subsection{End-to-End Performance}
\label{sec:evo_headline}

\begin{figure}[t]
\centering
\includegraphics[width=\columnwidth]{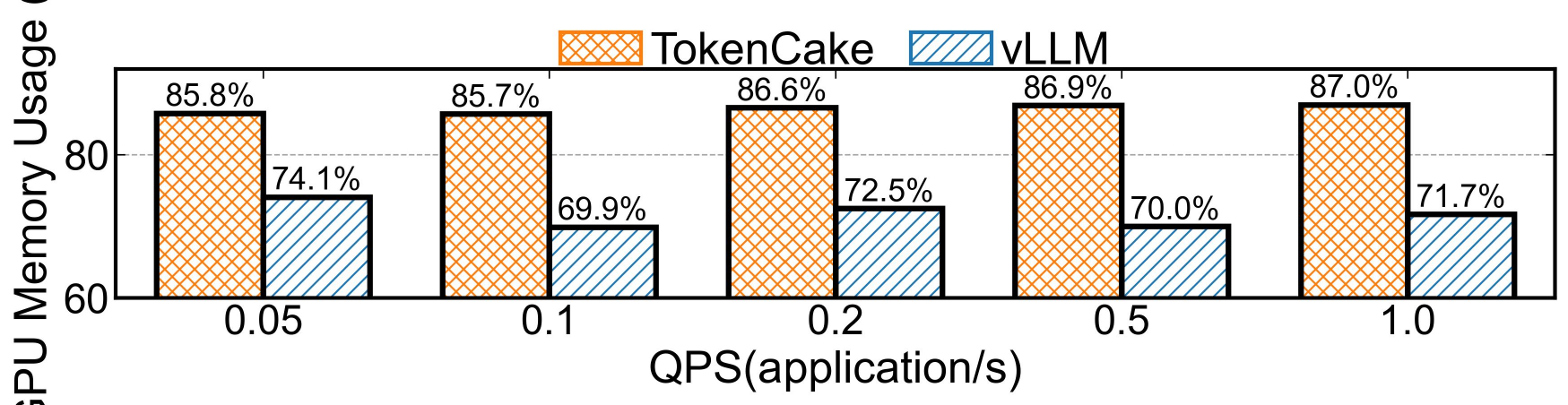}
\caption{
GPU \KVC utilization on Qwen2.5-14B Code-Writer under varying load.
}
\Description{Bar chart comparing GPU memory usage percentage between TokenCake (85.7\%--87.0\%) and vLLM (69.9\%--74.1\%) across five QPS levels (0.05, 0.1, 0.2, 0.5, 1.0). TokenCake maintains consistently higher utilization with a gap of 11.7--17.1 percentage points.}
\label{fig:evo_mem}
\vspace{-5mm}
\end{figure}

\noindent\textbf{Latency Under Varying Load.}
Figure~\ref{fig:evo_qps_compare} shows average end-to-end latency as a function of request arrival rate across three model sizes (Qwen2.5-14B, 32B, 72B), two applications (Code-Writer, Deep-Research), and two datasets (D1, D2).
\project consistently achieves the lowest latency across all configurations.
At low load (0.05 QPS), \project and vLLM perform comparably because memory contention is minimal.
As the arrival rate increases, latency for vLLM and vLLM-Prefix grows steeply because both retain stalled agents' \KVCs in GPU memory, saturating capacity and forcing smaller batch sizes.
Mooncake's remote caching provides partial relief at moderate load but cannot match \project because it remains agent-agnostic.
\project's latency scales more gradually because the \temporalscheduler offloads idle \KVCs during \fc stalls, freeing blocks for active computation.
The advantage holds across model sizes: on Qwen2.5-72B Code-Writer D2 at 1.0 QPS, \project reduces latency by 20.7\% relative to vLLM.
The benefit is amplified under tighter memory budgets: with \texttt{gpu\_mem\_utilization}=0.5
(Figure~\ref{fig:evo_mem}), \project reduces average latency by up to 47.06\% compared to vLLM, as agent-aware scheduling and offload coordination
reclaim GPU memory that other systems leave stranded on stalled agents.
Parrot is excluded from Figure~\ref{fig:evo_qps_compare} because its latency is 6.5--9.0$\times$ higher than \project's (Figure~\ref{fig:parrot_compare}); including it would compress the other curves and obscure the differences among competitive baselines.

\noindent\textbf{GPU Memory Utilization.}
The latency reduction is a direct consequence of better memory management.
Figure~\ref{fig:evo_mem} shows GPU \KVC utilization on Qwen2.5-14B Code-Writer.
\project maintains utilization at 85.7--87.0\% across all load levels, compared to 69.9--74.1\% for vLLM, an improvement of up to 16.9 percentage points.
The key difference is not total memory consumption but the utilization: \project keeps occupied blocks allocated to active, computation-ready requests, whereas vLLM's blocks are partly held by idle \KVCs from stalled agents that block new requests from being scheduled.
The same pattern appears on Qwen2.5-32B Code-Writer D2, where \project raises utilization from 53.5\% (vLLM) to 79.6\% at 1.0 QPS.

\subsection{Component Analysis}
\label{sec:component_controlled_studies}

We isolate the contribution of each component using four modes: \textit{baseline} (vLLM), \textit{agent} (only the \spatialscheduler), \textit{offload} (only the \temporalscheduler without agent awareness), and full \project (both).

\noindent\textbf{Scheduling and Offload Must Be Coordinated.}
On Code-Writer with Qwen2.5-14B (20 applications, 1.0 QPS, 0.5 GPU memory utilization), agent-aware scheduling alone reduces total latency from 502.2\,s to 424.8\,s (15.4\%).
Plain offload without agent awareness reduces total latency to 403.1\,s but performs 11{,}339 offloads---more than double the swap volume of full \project---and its average and P90 latencies (392.4\,s, 397.1\,s) remain close to baseline, indicating that indiscriminate offloading creates migration churn that delays some requests.
\project achieves the lowest latency across all metrics (344.6\,s total, 313.7\,s average, 328.3\,s P90) while reducing swap volume by 51\% compared to \textit{offload} alone.
Agent awareness allows the \temporalscheduler to target offloads at non-critical stalled agents and protect critical-path caches, achieving better results with fewer migrations.

\begin{figure}[t]
\centering
\includegraphics[width=\columnwidth]{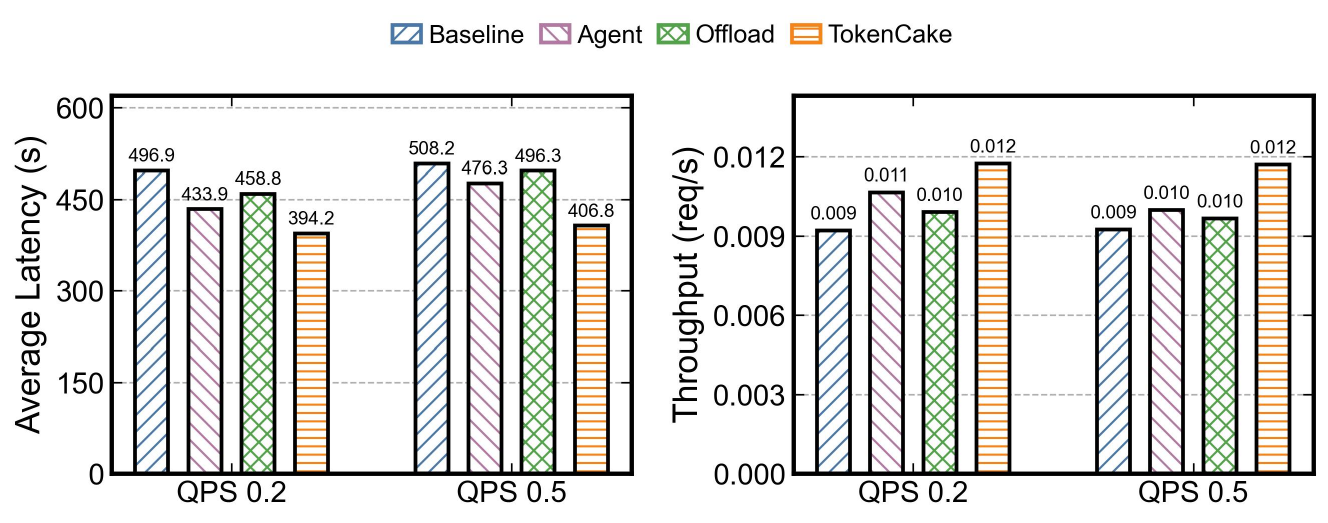}
\vspace{-6mm}
\caption{
Component behavior at 0.2 and 0.5 QPS (Qwen2.5-14B Code-Writer, 20 apps, 0.5 GPU memory utilization).
}
\Description{Dual-panel bar chart. Left: average latency for Baseline (496.9s, 508.2s), Agent (433.9s, 476.3s), Offload (458.8s, 496.3s), and TokenCake (394.2s, 406.8s) at QPS 0.2 and 0.5. Right: throughput where TokenCake achieves the highest values (0.012 req/s at both load points). TokenCake combines both components for the best result.}
\label{fig:ablation}
\vspace{-6mm}
\end{figure}

\noindent\textbf{Load Dependence.}
Figure~\ref{fig:ablation} shows component behavior at 0.2 and 0.5 QPS.
At both load points, \project achieves the lowest average latency and highest throughput.
At 0.2 QPS, \project reduces average latency from 496.9\,s to 394.2\,s (20.7\%); at 0.5 QPS, it reduces latency to 406.8\,s from 508.2\,s (19.9\%) and improves throughput by 26.6\%.
Across both load points, \textit{agent} alone consistently outperforms \textit{offload} alone, confirming that agent-aware scheduling provides a stronger standalone improvement than indiscriminate offloading.
Combining both mechanisms yields a further reduction that neither achieves independently.

\subsection{Remote-\KVC and Agent-Aware Baselines}
\label{sec:evo_remote_agent_baselines}

We compare \project with Mooncake and Parrot, which represent the two sides of the design space that \project bridges: \KVCc but agent-agnostic (Mooncake) and agent-aware but compute-centric (Parrot).

\begin{figure}[t]
\centering
\includegraphics[width=\columnwidth]{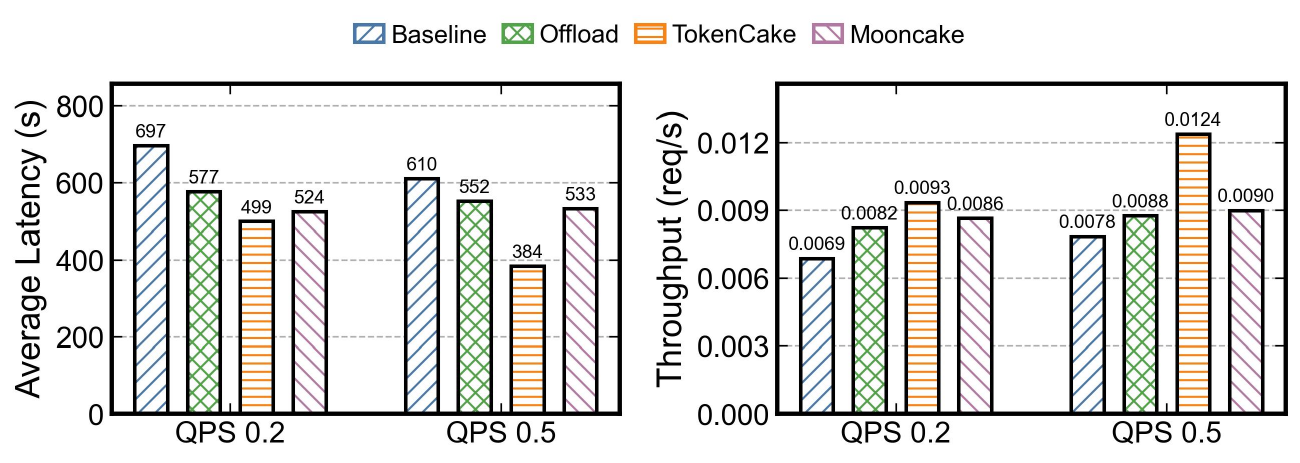}
\vspace{-6mm}
\caption{
Mooncake comparison on Qwen2.5-7B Code-Writer with five applications.
}
\Description{Dual-panel bar chart comparing Baseline, Offload, TokenCake, and Mooncake at QPS 0.2 and 0.5. Left: average latency where TokenCake achieves 499s and 384s vs. Mooncake's 524s and 533s. Right: throughput where TokenCake reaches 0.0124 req/s at QPS 0.5, 37.8\% higher than Mooncake's 0.0090 req/s.}
\label{fig:mooncake_compare}
\vspace{-5mm}
\end{figure}

\noindent\textbf{Mooncake.}
We connect Mooncake through the vLLM V1 Mooncake store connector with \textit{kv\_both} semantics.
To stress-test under limited GPU memory, we set \texttt{gpu\_mem\_utilization}=0.5 for this comparison;
all other end-to-end experiments in \S\ref{sec:evo_headline} use the default 0.8.
Figure~\ref{fig:mooncake_compare} compares Baseline (vLLM), Offload (\temporalscheduler only), \project, and Mooncake at two arrival rates.
At 0.2 QPS, Mooncake reduces average latency from 697\,s to 524\,s (24.8\%) and increases throughput from 0.0069 to 0.0086\,req/s, confirming that remote \KVC reuse provides a meaningful benefit.
\project reduces latency further to 499\,s, a 4.8\% improvement over Mooncake.
The modest gap at 0.2 QPS reflects low memory pressure: both systems have sufficient GPU blocks.
At 0.5 QPS, where memory contention intensifies, the gap widens: Mooncake reaches 533\,s while \project achieves 384\,s, a 28.0\% reduction over Mooncake and 37.0\% over Baseline.
Throughput follows the same pattern: \project reaches 0.0124\,req/s, a 37.8\% improvement over Mooncake's 0.0090\,req/s.
Offload alone (without agent awareness) performs worse than Mooncake at both load levels (577\,s vs.\ 524\,s at 0.2 QPS; 552\,s vs.\ 533\,s at 0.5 QPS), reinforcing the finding from \S\ref{sec:component_controlled_studies} that indiscriminate offloading introduces migration churn.
Only when offload is combined with agent-aware scheduling does \project surpass Mooncake, because the \spatialscheduler directs freed blocks to critical-path agents.
The fundamental difference is that Mooncake remains agent-agnostic: it does not use \fc events to predict idle intervals, nor does it prioritize caches based on graph criticality.

\begin{figure}[t]
\centering
\includegraphics[width=\columnwidth]{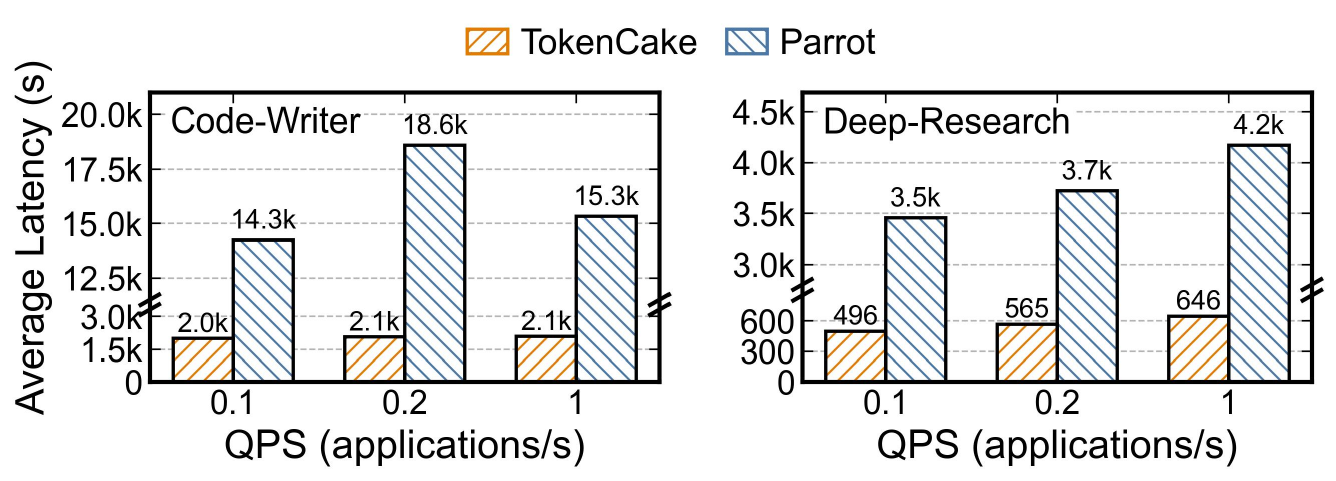}
\vspace{-6mm}
\caption{
Parrot comparison across three QPS levels on Code-Writer (left) and Deep-Research (right).
}
\Description{Dual-panel bar chart with broken y-axes. Left (Code-Writer): Parrot latency ranges from 14.3k--18.6k seconds while TokenCake achieves 2.0--2.1k seconds, a 7.1--9.0x gap. Right (Deep-Research): Parrot ranges from 3.5k--4.2k seconds while TokenCake achieves 496--646 seconds, a 6.5--7.0x gap across three QPS levels (0.1, 0.2, 1.0).}
\label{fig:parrot_compare}
\vspace{-5mm}
\end{figure}

\noindent\textbf{Parrot.}
Figure~\ref{fig:parrot_compare} compares \project with Parrot on both applications across three arrival rates (0.1, 0.2, 1.0 QPS) using D2.
On Code-Writer, Parrot's average latency ranges from 14.3k\,s to 18.6k\,s, while \project achieves 2.0--2.1k\,s, a gap of 7.1--9.0$\times$.
On Deep-Research, Parrot reports 3.5--4.2k\,s compared to \project's 496--646\,s, a gap of 6.5--7.0$\times$.
The gap persists across all load levels because the root cause is structural: Parrot does not manage \KVC memory, so non-critical agents exhaust GPU blocks and cause eviction of critical agents' caches regardless of scheduling order.
This comparison uses a different runtime (Parrot's own engine), so it serves as a system-scope check rather than a controlled experiment.
The result therefore shows an end-to-end system gap but does not isolate how much of that gap comes from \KVC management rather than differences between the two execution engines.

\subsection{Sensitivity Analysis}
\label{sec:evo_sensitivity}

We evaluate the \temporalscheduler's sensitivity to two factors: tool-time prediction error and policy parameter choices.
All experiments in this subsection use Qwen2.5-14B on a single A100-80\,GB PCIe GPU with \texttt{gpu\_mem\_utilization}=0.5.

\begin{figure}[t]
\centering
\includegraphics[width=\columnwidth]{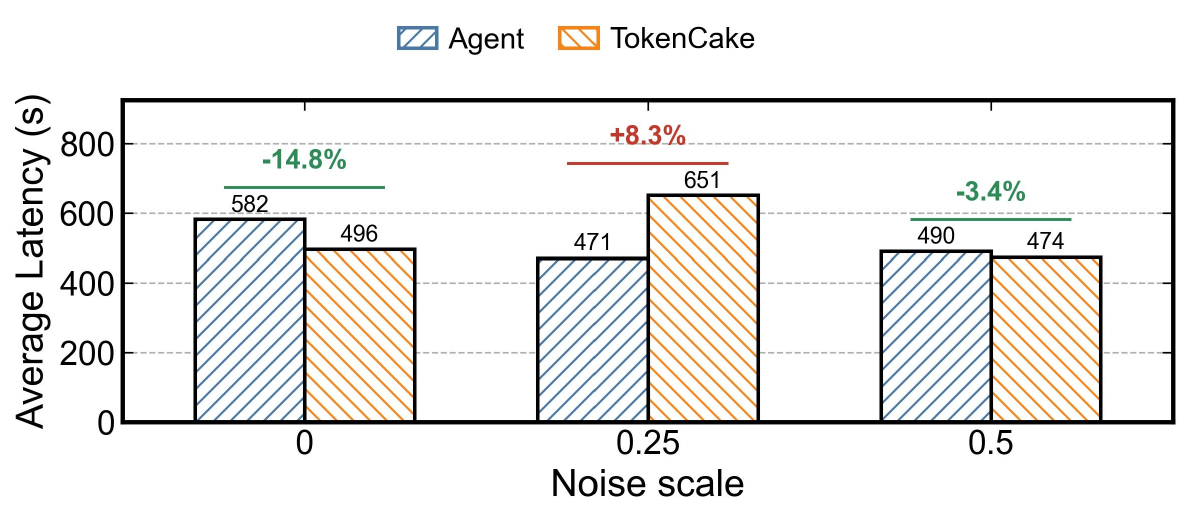}
\vspace{-9mm}
\caption{
Average latency of \textit{agent}-only and \project under tool-time noise; labels give the relative change of \project (negative is faster).
}
\Description{Bar chart comparing Agent-only and TokenCake average latency under three noise scales (0, 0.25, 0.5). At noise 0, TokenCake is 14.8\% faster (582s to 496s). At noise 0.25, TokenCake regresses by 8.3\% (471s to 651s). At noise 0.5, TokenCake recovers a 3.4\% improvement (490s to 474s), showing a non-monotonic pattern.}
\label{fig:noise}
\vspace{-5mm}
\end{figure}

\noindent\textbf{Tool-Time Variability.}
We inject multiplicative noise into \fc durations: at noise scale $s$, actual execution time is drawn from $[t \cdot (1{-}s),\; t \cdot (1{+}s)]$.
Figure~\ref{fig:noise} shows the latency change of \project relative to \textit{agent}-only scheduling.
With zero noise, \project reduces latency by 14.8\%.
At noise 0.25, \project regresses by 8.3\% because mistimed uploads pass the gate's feasibility check but still cause harmful migrations.
At noise 0.5, \project recovers a 3.4\% improvement because the larger errors cause scheduling windows to fail feasibility checks outright, so the gate correctly blocks most migrations.
This non-monotonic pattern reveals that intermediate prediction error is the hardest regime: hard rejections protect against large errors, but marginal errors that pass the gate impose disproportionate cost.
Applications with highly variable tool latencies should provide wider confidence intervals so the gate can apply a larger safety margin.

\begin{figure}[t]
\centering
\includegraphics[width=\columnwidth]{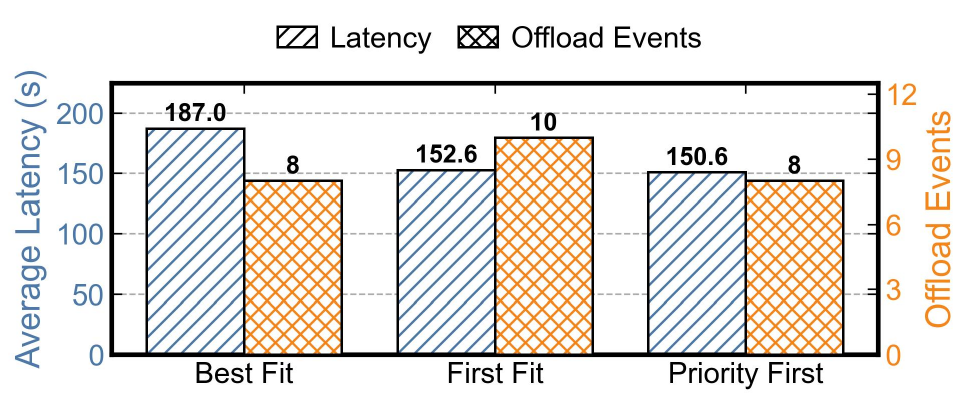}
\vspace{-9mm}
\caption{
Average latency (left axis) and offload event count (right axis) for three request-selection policies.
}
\Description{Dual-axis bar chart comparing Best Fit (187.0s, 8 offloads), First Fit (152.6s, 10 offloads), and Priority First (150.6s, 8 offloads). First Fit achieves the best overall tradeoff with low latency and high throughput while preserving the queue order optimized by the Spatial Scheduler.}
\label{fig:temporal_req_choose}
\end{figure}

\noindent\textbf{Temporal Request Selection.}
We vary two policy choices: the request-selection strategy used by the opportunistic gate and the spatial pressure watermark.
As shown in Figure~\ref{fig:temporal_req_choose}, for request selection, \textit{best\_fit} (select the request whose demand best matches freed capacity) performs worst (187.0\,s) because it reorders the queue and disrupts the \spatialscheduler's priorities.
\textit{priority\_first} (select the highest-priority request) achieves the lowest mean latency (150.6\,s) but inflates the tail (P95: 173.2\,s) by skipping small requests that would complete quickly.
\textit{first\_fit} (select the first queue-order request that fits) achieves the best overall tradeoff: 152.6\,s average, 164.7\,s P95, and the highest throughput (0.058\,req/s), because it preserves the queue order the \spatialscheduler has already optimized.
We use \textit{first\_fit} as the default.

\begin{figure}[t]
\centering
\includegraphics[width=\linewidth]{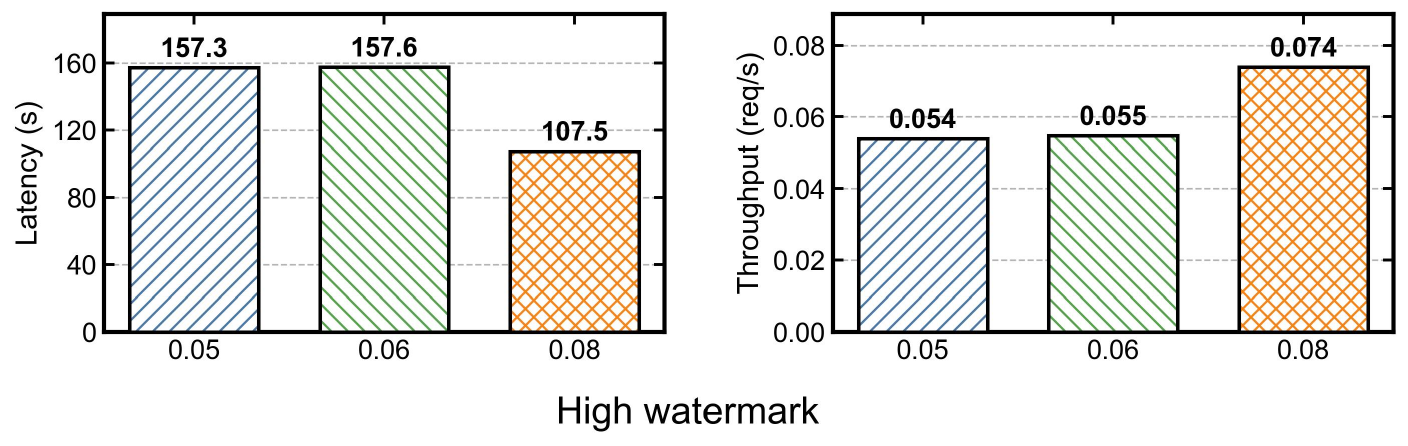}
\vspace{-9mm}
\caption{
Sensitivity to spatial pressure thresholds.
}
\Description{Dual-panel bar chart for high watermark thresholds 0.05, 0.06, and 0.08. Left: latency of 157.3s, 157.6s, and 107.5s respectively. Right: throughput of 0.054, 0.055, and 0.074 req/s. The 0.08 threshold achieves a 32\% latency reduction by correctly rejecting unnecessary offloads.}
\label{fig:spatial_knob}
\vspace{-5mm}
\end{figure}

\noindent\textbf{Spatial Pressure Thresholds.}
For the spatial pressure watermark (Figure~\ref{fig:spatial_knob}), aggressive thresholds (0.05, 0.06) trigger offloads frequently and produce similar latencies ($\sim$157\,s).
A higher watermark (0.08) rejects all offload candidates at this load level, achieving 107.5\,s---a 32\% reduction---because the gate correctly identifies that no waiting request can benefit from freed blocks.
This does not mean zero offload is universally optimal; it confirms the selectivity principle that offload should be conditional on whether freed blocks can admit useful active work.

\noindent\textbf{Hardware Dependence.}
The sensitivity effects above rely on the offload mechanism being actively triggered under sufficient memory pressure.
On hardware with higher memory bandwidth (e.g., H100), reduced transfer overhead causes the gate to reject most offload candidates, narrowing the differences between configurations.
Practitioners should therefore validate threshold choices on their target platform.

\subsection{Offload Overhead and Practicality}
\label{sec:evo_offload_tradeoff}

\begin{figure}[t]
\centering
\includegraphics[width=\columnwidth]{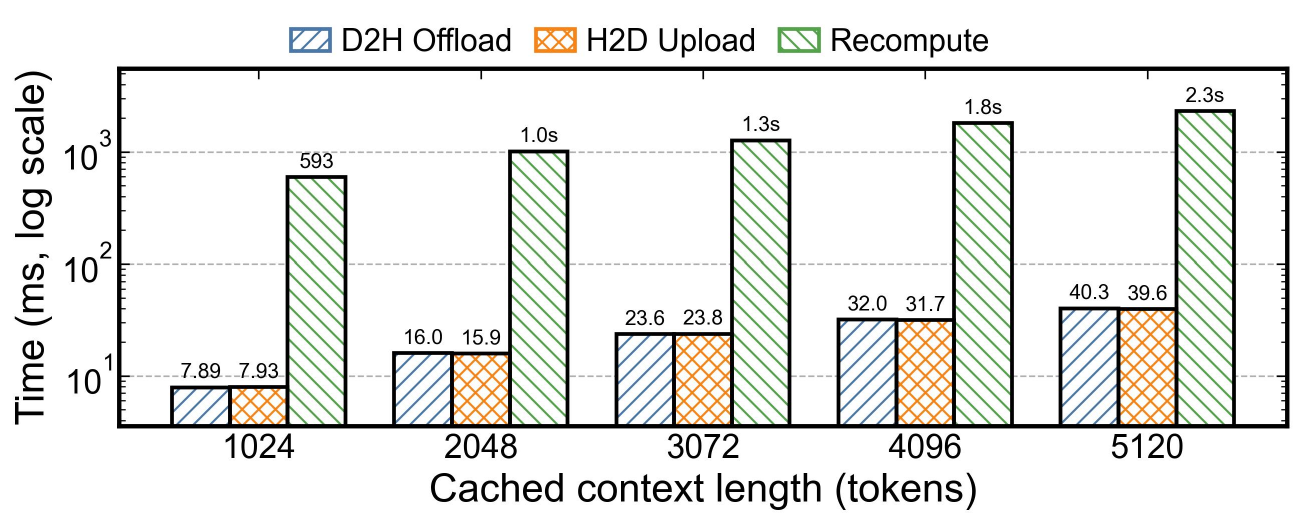}
\caption{
D2H offload, H2D upload, and recomputation time for Qwen2.5-14B on A100 PCIe.
}
\Description{Bar chart on log-scale y-axis (ms) comparing D2H Offload (7.89--40.3ms), H2D Upload (7.93--39.6ms), and Recompute (593ms--2.3s) times across cached context lengths from 1024 to 5120 tokens. At 4096 tokens, the round-trip migration (63.7ms) is 28.5x faster than recomputation (1.8s).}
\label{fig:evo_offload_time}
\end{figure}

\project assumes that preserving a stalled agent's \KVC in CPU memory is cheaper than discarding and recomputing it.
Figure~\ref{fig:evo_offload_time} validates this assumption on the A100 PCIe platform used for Qwen2.5-14B.
We measure cached context lengths from 1{,}024 to 5{,}120 tokens (64 to 320 blocks at 16 tokens/block, 3\,MiB/block in bfloat16).
At 4{,}096 tokens (256 blocks), offload takes 32.0\,ms and upload takes 31.7\,ms, for a 63.7\,ms round trip.
Recomputing the same context takes 1{,}815\,ms, making recomputation $28.5\times$ slower.
Across all measured lengths, recomputation is $26.8$--$37.5\times$ slower than a round-trip migration.

However, the round-trip cost still ranges from 15.8 to 79.8\,ms, and under high concurrency unnecessary migrations can consume PCIe bandwidth and occupy GPU blocks that active requests need.
This is why \project uses the opportunistic gate rather than an always-offload policy.

\section{Related Work}
\label{sec:related_work}

\noindent\textbf{LLM Application-Aware Scheduling.}
Parrot~\cite{parrot}, Agentix~\cite{agentix}, and Hermes~\cite{hermes} model applications as dependency graphs to mitigate head-of-line blocking;
Ayo~\cite{teola} pipelines LLM and tool stages to overlap execution.
These are compute-centric and do not manage \KVC memory, so non-critical agents can still evict critical-path caches.
Continuum~\cite{continuum} and ThunderAgent~\cite{thunderagent} retain agent state across tool calls but lack predictive migration and DAG-guided reservation.
VTC~\cite{fairness} ensures work-conserving fairness but neither models dependencies nor controls \KVC allocation.

\noindent\textbf{\KVC Memory Management.}
PagedAttention~\cite{vllm} and SGLang~\cite{sglang} eliminate fragmentation and reuse shared prefixes at the block level.
Mooncake~\cite{mooncake} and CachedAttention~\cite{cachedattention} offload \KVCs to host or disaggregated memory;
CacheBlend~\cite{cacheblend,lmcache} extends effective capacity through chunk-level reuse.
Other works reduce CPU-resident access cost~\cite{infinigen,cachegen,impress,directkv}, preserve session state or overlap I/O with computation~\cite{pensieve,hcache,neo,multiturndialogue}, improve eviction or load balancing~\cite{preble,kvcachewild}, exploit inter-agent segment reuse~\cite{cacheslide,tokendance}, or optimize compaction and sparse reconstruction~\cite{adaptivekvcache,diffkv,jenga,shadowkv}.
KVFlow~\cite{kvflow} is the closest workflow-aware manager: it prefetches CPU-resident prefixes using an Agent Step Graph, but does not use \fc boundaries to reclaim stalled \KVCs or reserve GPU blocks for critical-path work.
Orthogonally, prior work reduces \KVC footprint via prompt compression~\cite{llmlingua2,longllmlingua,coact} or manages expert-weight offloading in MoE models~\cite{pregatedmoe,finemoe,remoe}.

\noindent\textbf{Positioning.}
As summarized in Table~\ref{tab:kvc_offload_comparison}, agent-aware systems lack \KVC memory management, and \KVCc systems lack application context.
\project occupies the missing point by co-optimizing temporal \KVC lifecycle management and spatial memory reservation using the dependency graph and \fc events exposed by multi-agent applications.

\section{Discussion}
\label{sec:discussion}

\noindent\textbf{Multi-GPU Support and Scaling.}
The current evaluation includes a tensor-parallel configuration (Qwen2.5-72B on two NVIDIA H20 GPUs, \S\ref{sec:evo_headline}), demonstrating that \project's scheduling principles extend beyond a single device.
Scaling to larger GPU clusters opens additional opportunities: the \temporalscheduler's offload policy could target a neighboring GPU over NVLink as a faster destination than CPU memory, creating a tiered memory hierarchy, and the \spatialscheduler's reservation logic could coordinate across larger device pools to reduce cross-GPU critical inversion.
We are actively collaborating with the open-source community to extend \project's multi-GPU capabilities and look forward to broader deployment scenarios in future work.

\noindent\textbf{Dynamic Graphs.}
The current evaluation uses static application graphs for reproducibility.
In practice, some multi-agent applications contain dynamic edges where the LLM decides at runtime which downstream agent to invoke.
\project supports such runtime branching over branches declared in the DAG: skipped branches never enter the scheduler, and activated branches run as usual.
Within a fixed application, each agent type keeps a stable system prompt, graph role, and \KVC footprint, so the structural priority signals stay meaningful whichever branch runs.
The runtime priority terms and reservations adapt to the executed branches, because the \spatialscheduler periodically re-evaluates the critical agent set (\S\ref{sec:runtime_control}).
Integrating a probabilistic model of dynamic edges, as proposed in Hermes~\cite{hermes}, could improve reservation accuracy for applications with highly variable graph structure.

\noindent\textbf{Limitations.}
\project relies on developers to declare the agent DAG and tool types; unannotated calls fall back to a single opaque stage with a default time estimate.
The evaluation covers a single node with up to two tensor-parallel GPUs; cross-node deployment and a controlled comparison with Parrot on a common engine remain future work.

\section{Conclusion}
\label{sec:conclusion}

Multi-agent LLM applications with external \fcs expose temporal underutilization and spatial contention of \KVCs that existing systems address only in isolation.
\project exploits the observation that \fc events reveal both the idle interval and the criticality of each cache, and coordinates a \temporalscheduler that proactively offloads and predictively reloads idle \KVCs with a \spatialscheduler that reserves GPU capacity for critical-path agents.
On representative benchmarks, \project cuts end-to-end latency by up to 47.06\% in memory-constrained settings and improves GPU utilization by 16.9 percentage points over vLLM under high load, with ablations confirming that both schedulers are necessary.

\begin{acks}
We thank the anonymous reviewers and our shepherd for their valuable feedback.
We also thank the vLLM and Mooncake communities for their open-source contributions, and Kimi.
This work is supported by National Key R\&D Program of China under Grant No. 2024YFB4504300 and the Beijing Tongming Lake Information Technology Application Innovation Center (TLAIC).
Youwei Zhuo is the corresponding author.
\end{acks}

\balance

\bibliography{references}

@inproceedings{teola,
  title        = {Towards End-to-End Optimization of {LLM}-based Applications with {Ayo}},
  author       = {Xin Tan and Yimin Jiang and Yitao Yang and Hong Xu},
  booktitle    = {Proceedings of the 30th ACM International Conference on Architectural Support for Programming Languages and Operating Systems, Volume 2},
  series       = {ASPLOS '25},
  pages        = {1302--1316},
  year         = {2025},
  publisher    = {Association for Computing Machinery},
  address      = {New York, NY, USA},
  doi          = {10.1145/3676641.3716278},
  url          = {https://doi.org/10.1145/3676641.3716278},
  note         = {Preprint title: Teola}
}

@misc{hermes,
      title={Efficient Serving of LLM Applications with Probabilistic Demand Modeling},
      author={Yifei Liu and Zuo Gan and Zhenghao Gan and Weiye Wang and Chen Chen and Yizhou Shan and Xusheng Chen and Zhenhua Han and Yifei Zhu and Shixuan Sun and Minyi Guo},
      year={2025},
      eprint={2506.14851},
      archivePrefix={arXiv},
      primaryClass={cs.DC},
      url={https://arxiv.org/abs/2506.14851},
}

@inproceedings{cacheblend,
  author = {Yao, Jiayi and Li, Hanchen and Liu, Yuhan and Ray, Siddhant and Cheng, Yihua and Zhang, Qizheng and Du, Kuntai and Lu, Shan and Jiang, Junchen},
  title = {CacheBlend: Fast Large Language Model Serving for RAG with Cached Knowledge Fusion},
  year = {2025},
  url = {https://doi.org/10.1145/3689031.3696098},
  doi = {10.1145/3689031.3696098},
  booktitle = {Proceedings of the Twentieth European Conference on Computer Systems},
  pages = {94--109},
}

@inproceedings{parrot,
    author = {Chaofan Lin and Zhenhua Han and Chengruidong Zhang and Yuqing Yang and Fan Yang and Chen Chen and Lili Qiu},
    title = {Parrot: Efficient Serving of LLM-based Applications with Semantic Variable},
    booktitle = {18th USENIX Symposium on Operating Systems Design and Implementation (OSDI 24)},
    year = {2024},
    address = {Santa Clara, CA},
    publisher = {USENIX Association},
    url = {https://www.usenix.org/conference/osdi24/presentation/lin-chaofan},
    month = jul
}

@inproceedings{infercept,
  title={{InferCept}: Efficient Intercept Support for Augmented Large Language Model Inference},
  author={Abhyankar, Reyna and He, Zijian and Srivatsa, Vikranth and Zhang, Hao and Zhang, Yiying},
  booktitle={Proceedings of the 41st International Conference on Machine Learning},
  series={Proceedings of Machine Learning Research},
  volume={235},
  pages={81--95},
  year={2024},
  publisher={PMLR},
  url={https://proceedings.mlr.press/v235/abhyankar24a.html}
}

@misc{fc,
      title={Tool Calling: Enhancing Medication Consultation via Retrieval-Augmented Large Language Models}, 
      author={Zhongzhen Huang and Kui Xue and Yongqi Fan and Linjie Mu and Ruoyu Liu and Tong Ruan and Shaoting Zhang and Xiaofan Zhang},
      year={2024},
      eprint={2404.17897},
      archivePrefix={arXiv},
      primaryClass={cs.CL},
      url={https://arxiv.org/abs/2404.17897}, 
}

@misc{metagpt,
      title={MetaGPT: Meta Programming for A Multi-Agent Collaborative Framework}, 
      author={Sirui Hong and Mingchen Zhuge and Jiaqi Chen and Xiawu Zheng and Yuheng Cheng and Ceyao Zhang and Jinlin Wang and Zili Wang and Steven Ka Shing Yau and Zijuan Lin and Liyang Zhou and Chenyu Ran and Lingfeng Xiao and Chenglin Wu and Jürgen Schmidhuber},
      year={2024},
      eprint={2308.00352},
      archivePrefix={arXiv},
      primaryClass={cs.AI},
      url={https://arxiv.org/abs/2308.00352}, 
}

@misc{chatdev,
      title={ChatDev: Communicative Agents for Software Development}, 
      author={Chen Qian and Wei Liu and Hongzhang Liu and Nuo Chen and Yufan Dang and Jiahao Li and Cheng Yang and Weize Chen and Yusheng Su and Xin Cong and Juyuan Xu and Dahai Li and Zhiyuan Liu and Maosong Sun},
      year={2024},
      eprint={2307.07924},
      archivePrefix={arXiv},
      primaryClass={cs.SE},
      url={https://arxiv.org/abs/2307.07924}, 
}

@misc{autogen,
      title={AutoGen: Enabling Next-Gen LLM Applications via Multi-Agent Conversation}, 
      author={Qingyun Wu and Gagan Bansal and Jieyu Zhang and Yiran Wu and Beibin Li and Erkang Zhu and Li Jiang and Xiaoyun Zhang and Shaokun Zhang and Jiale Liu and Ahmed Hassan Awadallah and Ryen W White and Doug Burger and Chi Wang},
      year={2023},
      eprint={2308.08155},
      archivePrefix={arXiv},
      primaryClass={cs.AI},
      url={https://arxiv.org/abs/2308.08155}, 
}

@misc{tradingagents,
      title={TradingAgents: Multi-Agents LLM Financial Trading Framework}, 
      author={Yijia Xiao and Edward Sun and Di Luo and Wei Wang},
      year={2025},
      eprint={2412.20138},
      archivePrefix={arXiv},
      primaryClass={q-fin.TR},
      url={https://arxiv.org/abs/2412.20138}, 
}

@misc{agentsociety,
      title={AgentSociety: Large-Scale Simulation of LLM-Driven Generative Agents Advances Understanding of Human Behaviors and Society}, 
      author={Jinghua Piao and Yuwei Yan and Jun Zhang and Nian Li and Junbo Yan and Xiaochong Lan and Zhihong Lu and Zhiheng Zheng and Jing Yi Wang and Di Zhou and Chen Gao and Fengli Xu and Fang Zhang and Ke Rong and Jun Su and Yong Li},
      year={2025},
      eprint={2502.08691},
      archivePrefix={arXiv},
      primaryClass={cs.SI},
      url={https://arxiv.org/abs/2502.08691}, 
}

@inproceedings{vllm,
author = {Kwon, Woosuk and Li, Zhuohan and Zhuang, Siyuan and Sheng, Ying and Zheng, Lianmin and Yu, Cody Hao and Gonzalez, Joseph and Zhang, Hao and Stoica, Ion},
title = {Efficient Memory Management for Large Language Model Serving with PagedAttention},
year = {2023},
isbn = {9798400702297},
publisher = {Association for Computing Machinery},
address = {New York, NY, USA},
url = {https://doi.org/10.1145/3600006.3613165},
doi = {10.1145/3600006.3613165},
booktitle = {Proceedings of the 29th Symposium on Operating Systems Principles},
pages = {611--626},
numpages = {16},
location = {Koblenz, Germany},
series = {SOSP '23}
}

@inproceedings{sglang,
  title={SGLang: Efficient Execution of Structured Language Model Programs},
  author={Lianmin Zheng and Liangsheng Yin and Zhiqiang Xie and Chuyue Sun and Jeff Huang and Cody Hao Yu and Shiyi Cao and Christos Kozyrakis and Ion Stoica and Joseph E. Gonzalez and Clark Barrett and Ying Sheng},
  booktitle={Advances in Neural Information Processing Systems},
  volume={37},
  pages={62557--62583},
  year={2024},
  publisher={Curran Associates, Inc.},
  url={https://proceedings.neurips.cc/paper_files/paper/2024/hash/724be4472168f31ba1c9ac630f15dec8-Abstract-Conference.html}
}

@inproceedings{agentix,
  title={Agentix: An Efficient Serving Engine for {LLM} Agents as General Programs},
  author={Michael Luo and Xiaoxiang Shi and Colin Cai and Tianjun Zhang and Justin Wong and Yichuan Wang and Chi Wang and Yanping Huang and Zhifeng Chen and Joseph E. Gonzalez and Ion Stoica},
  booktitle={23rd USENIX Symposium on Networked Systems Design and Implementation (NSDI 26)},
  pages={2443--2459},
  year={2026},
  publisher={USENIX Association},
  url={https://www.usenix.org/conference/nsdi26/presentation/luo},
  month=may
}

@inproceedings{mooncake,
  title={Mooncake: Trading More Storage for Less Computation---A {KVCache-centric} Architecture for Serving {LLM} Chatbot},
  author={Ruoyu Qin and Zheming Li and Weiran He and Jialei Cui and Feng Ren and Mingxing Zhang and Yongwei Wu and Weimin Zheng and Xinran Xu},
  booktitle={23rd USENIX Conference on File and Storage Technologies (FAST 25)},
  pages={155--170},
  year={2025},
  publisher={USENIX Association},
  url={https://www.usenix.org/conference/fast25/presentation/qin},
  month=feb
}

@online{mcp,
  title = {Model Context Protocol Specification (Version 2025-06-18)},
  author = {{Anthropic, Inc.}},
  url   = {https://modelcontextprotocol.io/specification/2025-06-18},
  year  = {2025},
  note  = {Accessed: 2026-09-17}
}

@misc{agentscope,
      title={AgentScope: A Flexible yet Robust Multi-Agent Platform}, 
      author={Dawei Gao and Zitao Li and Xuchen Pan and Weirui Kuang and Zhijian Ma and Bingchen Qian and Fei Wei and Wenhao Zhang and Yuexiang Xie and Daoyuan Chen and Liuyi Yao and Hongyi Peng and Zeyu Zhang and Lin Zhu and Chen Cheng and Hongzhu Shi and Yaliang Li and Bolin Ding and Jingren Zhou},
      year={2024},
      eprint={2402.14034},
      archivePrefix={arXiv},
      primaryClass={cs.MA},
      url={https://arxiv.org/abs/2402.14034}, 
}

@misc{sharegpt,
  title        = {ShareGPT Vicuna Unfiltered},
  author       = {anon8231489123},
  year         = {2023},
  url          = {https://huggingface.co/datasets/anon8231489123/ShareGPT_Vicuna_unfiltered},
  note         = {Hugging Face dataset},
}

@misc{agentcode,
  title        = {AgentCode: A dataset for code generated by LLM agents},
  author       = {AlignmentLab AI},
  year         = {2024},
  howpublished = {Hugging Face Datasets},
  url          = {https://huggingface.co/datasets/AlignmentLab-AI/agentcode},
  note         = {Accessed: 2025-09-01}
}

@inproceedings{cachedattention,
author = {Gao, Bin and He, Zhuomin and Sharma, Puru and Kang, Qingxuan and Jevdjic, Djordje and Deng, Junbo and Yang, Xingkun and Yu, Zhou and Zuo, Pengfei},
title = {Cost-efficient large language model serving for multi-turn conversations with CachedAttention},
year = {2024},
isbn = {978-1-939133-41-0},
publisher = {USENIX Association},
address = {USA},
booktitle = {Proceedings of the 2024 USENIX Conference on Usenix Annual Technical Conference},
articleno = {7},
numpages = {16},
location = {Santa Clara, CA, USA},
series = {USENIX ATC'24}
}

@misc{gemini-deep-research,
  author       = {Google Gemini Team},
  title        = {Gemini Fullstack LangGraph Quickstart},
  year         = {2025},
  howpublished = {\url{https://github.com/google-gemini/gemini-fullstack-langgraph-quickstart}},
  note         = {Accessed: 2025-09-23}
}

@misc{anthropic_claude_code,
  author       = {{Anthropic}},
  title        = {Enabling {Claude Code} to Work More Autonomously},
  howpublished = {\url{https://www.anthropic.com/news/enabling-claude-code-to-work-more-autonomously}},
  year         = {2025},
  month        = sep,
  note         = {Accessed: 2026-09-17}
}

@misc{openai_codex,
  author       = {{OpenAI}},
  title        = {Introducing the {Codex} App},
  howpublished = {\url{https://openai.com/index/introducing-the-codex-app/}},
  year         = {2026},
  month        = feb,
  note         = {Accessed: 2026-09-17}
}

@misc{google_gemini_deep_research,
  author       = {{Google}},
  title        = {Gemini {Deep Research}: Your Personal {AI} Research Assistant},
  howpublished = {\url{https://gemini.google/overview/deep-research/}},
  year         = {2024},
  note         = {Accessed: 2026-05-15}
}

@misc{openai_deep_research,
  author       = {{OpenAI}},
  title        = {Introducing Deep Research},
  howpublished = {\url{https://openai.com/index/introducing-deep-research/}},
  year         = {2025},
  month        = feb,
  note         = {Accessed: 2026-05-15}
}

@misc{anthropic_claude_research,
  author       = {{Anthropic}},
  title        = {Introducing {Claude} Research},
  howpublished = {\url{https://claude.com/blog/research}},
  year         = {2025},
  note         = {Accessed: 2026-05-15}
}

@article{coact,
  author       = {Haorui Chen and
                  Yuancheng Zhu and
                  Yitong Zhang and
                  Jia Li},
  title        = {CoACT: Action-Preserving Observation Compression for Coding Agents},
  journal      = {CoRR},
  volume       = {abs/2607.02911},
  year         = {2026},
  url          = {https://doi.org/10.48550/arXiv.2607.02911},
  doi          = {10.48550/ARXIV.2607.02911},
  eprinttype   = {arXiv},
  eprint       = {2607.02911},
}

@inproceedings{remoe,
  title        = {ReMoE: Boosting Expert Reuse through Router Fine-Tuning in Memory-Constrained {MoE} {LLM} Inference},
  author       = {Xiongwei Zhu and Xiaojian Liao and Tianyang Jiang and Yusen Zhang and Liang Wang and Limin Xiao},
  booktitle    = {Forty-third International Conference on Machine Learning},
  year         = {2026},
  url          = {https://openreview.net/forum?id=ylAhgNb2ak},
}

@inproceedings{llmlingua2,
  title={Llmlingua-2: Data distillation for efficient and faithful task-agnostic prompt compression},
  author={Pan, Zhuoshi and Wu, Qianhui and Jiang, Huiqiang and Xia, Menglin and Luo, Xufang and Zhang, Jue and Lin, Qingwei and R{\"u}hle, Victor and Yang, Yuqing and Lin, Chin-Yew and others},
  booktitle={Findings of the Association for Computational Linguistics: ACL 2024},
  pages={963--981},
  year={2024}
}

@inproceedings{longllmlingua,
  title={Longllmlingua: Accelerating and enhancing llms in long context scenarios via prompt compression},
  author={Jiang, Huiqiang and Wu, Qianhui and Luo, Xufang and Li, Dongsheng and Lin, Chin-Yew and Yang, Yuqing and Qiu, Lili},
  booktitle={Proceedings of the 62nd Annual Meeting of the Association for Computational Linguistics (Volume 1: Long Papers)},
  pages={1658--1677},
  year={2024}
}

@inproceedings{pregatedmoe,
  title={Pre-gated moe: An algorithm-system co-design for fast and scalable mixture-of-expert inference},
  author={Hwang, Ranggi and Wei, Jianyu and Cao, Shijie and Hwang, Changho and Tang, Xiaohu and Cao, Ting and Yang, Mao},
  booktitle={2024 ACM/IEEE 51st Annual International Symposium on Computer Architecture (ISCA)},
  pages={1018--1031},
  year={2024},
  organization={IEEE}
}

@inproceedings{finemoe,
  title={Taming latency-memory trade-off in MoE-based LLM serving via fine-grained expert offloading},
  author={Yu, Hanfei and Cui, Xingqi and Zhang, Hong and Wang, Hao and Wang, Hao},
  booktitle={Proceedings of the 21st European Conference on Computer Systems},
  pages={176--191},
  year={2026}
}

@inproceedings{fairness,
  title={Fairness in Serving Large Language Models},
  author={Ying Sheng and Shiyi Cao and Dacheng Li and Banghua Zhu and Zhuohan Li and Danyang Zhuo and Joseph E. Gonzalez and Ion Stoica},
  booktitle={18th USENIX Symposium on Operating Systems Design and Implementation (OSDI 24)},
  pages={965--988},
  year={2024},
  publisher={USENIX Association},
  url={https://www.usenix.org/conference/osdi24/presentation/sheng},
  month=jul
}

@inproceedings{infinigen,
  title={{InfiniGen}: Efficient Generative Inference of Large Language Models with Dynamic {KV} Cache Management},
  author={Wonbeom Lee and Jungi Lee and Junghwan Seo and Jaewoong Sim},
  booktitle={18th USENIX Symposium on Operating Systems Design and Implementation (OSDI 24)},
  pages={155--172},
  year={2024},
  publisher={USENIX Association},
  url={https://www.usenix.org/conference/osdi24/presentation/lee},
  month=jul
}

@inproceedings{cachegen,
  title={CacheGen: {KV} Cache Compression and Streaming for Fast Large Language Model Serving},
  author={Yuhan Liu and Hanchen Li and Yihua Cheng and Siddhant Ray and Yuyang Huang and Qizheng Zhang and Kuntai Du and Jiayi Yao and Shan Lu and Ganesh Ananthanarayanan and Michael Maire and Henry Hoffmann and Ari Holtzman and Junchen Jiang},
  booktitle={Proceedings of the ACM SIGCOMM 2024 Conference},
  pages={38--56},
  year={2024},
  publisher={Association for Computing Machinery},
  doi={10.1145/3651890.3672274},
  url={https://doi.org/10.1145/3651890.3672274}
}

@inproceedings{impress,
  title={{IMPRESS}: An Importance-Informed Multi-Tier Prefix {KV} Storage System for Large Language Model Inference},
  author={Weijian Chen and Shuibing He and Haoyang Qu and Ruidong Zhang and Siling Yang and Ping Chen and Yi Zheng and Baoxing Huai and Gang Chen},
  booktitle={23rd USENIX Conference on File and Storage Technologies (FAST 25)},
  pages={187--201},
  year={2025},
  publisher={USENIX Association},
  url={https://www.usenix.org/conference/fast25/presentation/chen-weijian-impress},
  month=feb
}

@inproceedings{directkv,
  title={No Buffer, No Bottleneck: Efficient Zero-Copy {KV} Cache Offloading for Long-Context {LLMs}},
  author={Shutian Luo and Haiying Shen},
  booktitle={20th USENIX Symposium on Operating Systems Design and Implementation (OSDI 26)},
  pages={39--54},
  year={2026},
  publisher={USENIX Association},
  url={https://www.usenix.org/conference/osdi26/presentation/luo},
  month=jul
}

@inproceedings{cacheslide,
  title={{CacheSlide}: Unlocking Cross Position-Aware {KV} Cache Reuse for Accelerating {LLM} Serving},
  author={Yang Liu and Yunfei Gu and Liqiang Zhang and Chentao Wu and Guangtao Xue and Jie Li and Minyi Guo and Junhao Hu and Jie Meng},
  booktitle={24th USENIX Conference on File and Storage Technologies (FAST 26)},
  pages={83--99},
  year={2026},
  publisher={USENIX Association},
  url={https://www.usenix.org/conference/fast26/presentation/liu-yang},
  month=feb
}

@article{continuum,
  title={Continuum: Efficient and Robust Multi-Turn {LLM} Agent Scheduling with {KV} Cache Time-to-Live},
  author={Hanchen Li and Runyuan He and Qiuyang Mang and Qizheng Zhang and Huanzhi Mao and Xiaokun Chen and Hangrui Zhou and Huanchen Zhang and Alvin Cheung and Joseph E. Gonzalez and Ion Stoica},
  journal={arXiv preprint arXiv:2511.02230},
  year={2025},
  url={https://arxiv.org/abs/2511.02230}
}

@article{lmcache,
  title={{LMCache}: An Efficient {KV} Cache Layer for Enterprise-Scale {LLM} Inference},
  author={Yuhan Liu and Yihua Cheng and Jiayi Yao and Yuwei An and Xiaokun Chen and Shaoting Feng and Yuyang Huang and Samuel Shen and Rui Zhang and Kuntai Du and Junchen Jiang},
  journal={arXiv preprint arXiv:2510.09665},
  year={2025},
  url={https://arxiv.org/abs/2510.09665}
}

@inproceedings{sweagent,
  title={{SWE-agent}: Agent-Computer Interfaces Enable Automated Software Engineering},
  author={John Yang and Carlos E. Jimenez and Alexander Wettig and Kilian Lieret and Shunyu Yao and Karthik Narasimhan and Ofir Press},
  booktitle={Advances in Neural Information Processing Systems},
  volume={37},
  pages={50528--50652},
  year={2024},
  url={https://proceedings.neurips.cc/paper_files/paper/2024/hash/5a7c947568c1b1328ccc5230172e1e7c-Abstract-Conference.html}
}

@inproceedings{openhands,
  title={{OpenHands}: An Open Platform for {AI} Software Developers as Generalist Agents},
  author={Xingyao Wang and Boxuan Li and Yufan Song and Frank F. Xu and Xiangru Tang and Mingchen Zhuge and Jiayi Pan and Yueqi Song and Bowen Li and Jaskirat Singh and Hoang Tran and Fuqiang Li and Ren Ma and Mingzhang Zheng and Bill Qian and Daniel Shao and Niklas Muennighoff and Yizhe Zhang and Binyuan Hui and Junyang Lin and Robert Brennan and Hao Peng and Heng Ji and Graham Neubig},
  booktitle={International Conference on Learning Representations},
  year={2025},
  url={https://proceedings.iclr.cc/paper_files/paper/2025/hash/a4b6ad6b48850c0c331d1259fc66a69c-Abstract-Conference.html}
}

@inproceedings{swebench,
  title={{SWE-bench}: Can Language Models Resolve Real-World {GitHub} Issues?},
  author={Carlos E. Jimenez and John Yang and Alexander Wettig and Shunyu Yao and Kexin Pei and Ofir Press and Karthik Narasimhan},
  booktitle={International Conference on Learning Representations},
  year={2024},
  url={https://proceedings.iclr.cc/paper_files/paper/2024/hash/edac78c3e300629acfe6cbe9ca88fb84-Abstract-Conference.html}
}

@inproceedings{storm,
  title={Assisting in Writing Wikipedia-like Articles From Scratch with Large Language Models},
  author={Yijia Shao and Yucheng Jiang and Theodore Kanell and Peter Xu and Omar Khattab and Monica S. Lam},
  booktitle={Proceedings of the 2024 Conference of the North American Chapter of the Association for Computational Linguistics: Human Language Technologies},
  pages={6252--6278},
  year={2024},
  publisher={Association for Computational Linguistics},
  doi={10.18653/v1/2024.naacl-long.347},
  url={https://aclanthology.org/2024.naacl-long.347/}
}

@inproceedings{costorm,
  title={Into the Unknown Unknowns: Engaged Human Learning through Participation in Language Model Agent Conversations},
  author={Yucheng Jiang and Yijia Shao and Dekun Ma and Sina Semnani and Monica S. Lam},
  booktitle={Proceedings of the 2024 Conference on Empirical Methods in Natural Language Processing},
  pages={9917--9955},
  year={2024},
  publisher={Association for Computational Linguistics},
  doi={10.18653/v1/2024.emnlp-main.554},
  url={https://aclanthology.org/2024.emnlp-main.554/}
}

@inproceedings{webthinker,
  title={{WebThinker}: Empowering Large Reasoning Models with Deep Research Capability},
  author={Xiaoxi Li and Jiajie Jin and Guanting Dong and Hongjin Qian and Yongkang Wu and Ji-Rong Wen and Yutao Zhu and Zhicheng Dou},
  booktitle={Advances in Neural Information Processing Systems},
  volume={38},
  year={2025},
  url={https://proceedings.neurips.cc/paper_files/paper/2025/hash/ae03bdef276132fae089692445725635-Abstract-Conference.html}
}

@inproceedings{mind2web2,
  title={{Mind2Web 2}: Evaluating Agentic Search with Agent-as-a-Judge},
  author={Boyu Gou and Zanming Huang and Yuting Ning and Yu Gu and Michael Lin and Weijian Qi and Andrei Kopanev and Botao Yu and Bernal Jimenez Gutierrez and Yiheng Shu and Chan Hee Song and Jiaman Wu and Shijie Chen and Hanane Moussa and Tianshu Zhang and Jian Xie and Yifei Li and Tianci Xue and Zeyi Liao and Kai Zhang and Boyuan Zheng and Zhaowei Cai and Viktor Rozgic and Morteza Ziyadi and Huan Sun and Yu Su},
  booktitle={Advances in Neural Information Processing Systems},
  volume={38},
  year={2025},
  url={https://proceedings.neurips.cc/paper_files/paper/2025/hash/fdcec9f5b99aa4fc8f4fb8487802d737-Abstract-Datasets_and_Benchmarks_Track.html}
}

@inproceedings{gaia,
  title={{GAIA}: A Benchmark for General {AI} Assistants},
  author={Gregoire Mialon and Clementine Fourrier and Craig Swift and Thomas Wolf and Yann LeCun and Thomas Scialom},
  booktitle={International Conference on Learning Representations},
  year={2024},
  url={https://proceedings.iclr.cc/paper_files/paper/2024/hash/25ae35b5b1738d80f1f03a8713e405ec-Abstract-Conference.html}
}

@inproceedings{react,
  title={{ReAct}: Synergizing Reasoning and Acting in Language Models},
  author={Shunyu Yao and Jeffrey Zhao and Dian Yu and Nan Du and Izhak Shafran and Karthik Narasimhan and Yuan Cao},
  booktitle={International Conference on Learning Representations},
  year={2023},
  url={https://openreview.net/forum?id=WE_vluYUL-X}
}

@inproceedings{toolformer,
  title={Toolformer: Language Models Can Teach Themselves to Use Tools},
  author={Timo Schick and Jane Dwivedi-Yu and Roberto Dessi and Roberta Raileanu and Maria Lomeli and Eric Hambro and Luke Zettlemoyer and Nicola Cancedda and Thomas Scialom},
  booktitle={Advances in Neural Information Processing Systems},
  volume={36},
  pages={68539--68551},
  year={2023},
  url={https://proceedings.neurips.cc/paper/2023/hash/d842425e4bf79ba039352da0f658a906-Abstract-Conference.html}
}

@inproceedings{toolllm,
  title={{ToolLLM}: Facilitating Large Language Models to Master 16000+ Real-World {APIs}},
  author={Yujia Qin and Shihao Liang and Yining Ye and Kunlun Zhu and Lan Yan and Yaxi Lu and Yankai Lin and Xin Cong and Xiangru Tang and Bill Qian and Sihan Zhao and Lauren Hong and Runchu Tian and Ruobing Xie and Jie Zhou and Mark Gerstein and Dahai Li and Zhiyuan Liu and Maosong Sun},
  booktitle={International Conference on Learning Representations},
  year={2024},
  url={https://proceedings.iclr.cc/paper_files/paper/2024/hash/28e50ee5b72e90b50e7196fde8ea260e-Abstract-Conference.html}
}

@inproceedings{bfcl,
  title={The Berkeley Function Calling Leaderboard ({BFCL}): From Tool Use to Agentic Evaluation of Large Language Models},
  author={Shishir G. Patil and Huanzhi Mao and Fanjia Yan and Charlie Cheng-Jie Ji and Vishnu Suresh and Ion Stoica and Joseph E. Gonzalez},
  booktitle={Proceedings of the 42nd International Conference on Machine Learning},
  series={Proceedings of Machine Learning Research},
  volume={267},
  pages={48371--48392},
  year={2025},
  publisher={PMLR},
  url={https://proceedings.mlr.press/v267/patil25a.html}
}

@inproceedings{mcpmark,
  title={{MCPMark}: A Benchmark for Stress-Testing Realistic and Comprehensive {MCP} Use},
  author={Zijian Wu and Xiangyan Liu and Xinyuan Zhang and Lingjun Chen and Fanqing Meng and Lingxiao Du and Yiran Zhao and Fanshi Zhang and Yaoqi Ye and Jiawei Wang and Zirui Wang and Jinjie Ni and Yufan Yang and Arvin Xu and Michael Qizhe Shieh},
  booktitle={International Conference on Learning Representations},
  year={2026},
  url={https://proceedings.iclr.cc/paper_files/paper/2026/hash/8138d211ce8790fdfbeeeb9781838a37-Abstract-Conference.html}
}

@article{agentsysbench,
  title={From {LLM} Inference to Agentic Workloads: Characterization and Implications for Serving Systems},
  author={Chaokun Chang and Yukun Zhou and Kaihua Fu and Dakai An and Tianyu Feng and Hanfeng Lu and Sheng Yao and Pu Guo and Yinghao Yu and Yizhou Shan and Bo Li and Binhang Yuan and Wei Wang},
  journal={arXiv preprint arXiv:2608.15127},
  year={2026},
  url={https://arxiv.org/abs/2608.15127}
}

@inproceedings{thunderagent,
  title={ThunderAgent: A Simple, Fast and Program-Aware Agentic Inference System},
  author={Hao Kang and Ziyang Li and Weili Xu and Xinyu Yang and Yinfang Chen and Junxiong Wang and Beidi Chen and Tushar Krishna and Chenfeng Xu and Simran Arora},
  booktitle={Proceedings of the 43rd International Conference on Machine Learning},
  series={Proceedings of Machine Learning Research},
  volume={306},
  year={2026},
  publisher={PMLR},
  url={https://arxiv.org/abs/2602.13692}
}

@inproceedings{pensieve,
  title={Stateful Large Language Model Serving with Pensieve},
  author={Lingfan Yu and Jinkun Lin and Jinyang Li},
  booktitle={Proceedings of the Twentieth European Conference on Computer Systems},
  pages={144--158},
  year={2025},
  publisher={Association for Computing Machinery},
  doi={10.1145/3689031.3696086},
  url={https://doi.org/10.1145/3689031.3696086}
}

@inproceedings{hcache,
  title={Fast State Restoration in {LLM} Serving with {HCache}},
  author={Shiwei Gao and Youmin Chen and Jiwu Shu},
  booktitle={Proceedings of the Twentieth European Conference on Computer Systems},
  pages={128--143},
  year={2025},
  publisher={Association for Computing Machinery},
  doi={10.1145/3689031.3696072},
  url={https://doi.org/10.1145/3689031.3696072}
}

@inproceedings{multiturndialogue,
  title={Accelerating {LLM} Serving for Multi-turn Dialogues with Efficient Resource Management},
  author={Jinwoo Jeong and Jeongseob Ahn},
  booktitle={Proceedings of the 30th ACM International Conference on Architectural Support for Programming Languages and Operating Systems, Volume 2},
  pages={1--15},
  year={2025},
  publisher={Association for Computing Machinery},
  doi={10.1145/3676641.3716245},
  url={https://doi.org/10.1145/3676641.3716245}
}

@inproceedings{kvcachewild,
  title={{KVCache} Cache in the Wild: Characterizing and Optimizing {KVCache} Cache at a Large Cloud Provider},
  author={Jiahao Wang and Jinbo Han and Xingda Wei and Sijie Shen and Dingyan Zhang and Chenguang Fang and Rong Chen and Wenyuan Yu and Haibo Chen},
  booktitle={2025 USENIX Annual Technical Conference (USENIX ATC 25)},
  pages={465--482},
  year={2025},
  publisher={USENIX Association},
  url={https://www.usenix.org/conference/atc25/presentation/wang-jiahao},
  month=jul
}

@inproceedings{adaptivekvcache,
  title={High Throughput and Low Latency {LLM} Serving via Adaptive {KV} Caching},
  author={Wenyan Chen and Chengzhi Lu and Huanle Xu and Kejiang Ye and ChengZhong Xu},
  booktitle={Proceedings of the 21st European Conference on Computer Systems},
  pages={56--71},
  year={2026},
  publisher={Association for Computing Machinery},
  doi={10.1145/3767295.3803570},
  url={https://doi.org/10.1145/3767295.3803570}
}

@inproceedings{neo,
  title={{NEO}: Saving {GPU} Memory Crisis with {CPU} Offloading for Online {LLM} Inference},
  author={Xuanlin Jiang and Yang Zhou and Shiyi Cao and Ion Stoica and Minlan Yu},
  booktitle={Proceedings of Machine Learning and Systems},
  volume={7},
  year={2025},
  url={https://proceedings.mlsys.org/paper_files/paper/2025/hash/66a026c0d17040889b50f0dfa650e5e0-Abstract-Conference.html}
}

@inproceedings{preble,
  title={Preble: Efficient Distributed Prompt Scheduling for {LLM} Serving},
  author={Vikranth Srivatsa and Zijian He and Reyna Abhyankar and Dongming Li and Yiying Zhang},
  booktitle={International Conference on Learning Representations},
  year={2025},
  url={https://proceedings.iclr.cc/paper_files/paper/2025/hash/5bc342f48de8264779952fac378f96dc-Abstract-Conference.html}
}

@inproceedings{diffkv,
  title={{DiffKV}: Differentiated Memory Management for Large Language Models with Parallel {KV} Compaction},
  author={Yanqi Zhang and Yuwei Hu and Runyuan Zhao and John C. S. Lui and Haibo Chen},
  booktitle={Proceedings of the ACM SIGOPS 31st Symposium on Operating Systems Principles},
  pages={431--445},
  year={2025},
  publisher={Association for Computing Machinery},
  doi={10.1145/3731569.3764810},
  url={https://doi.org/10.1145/3731569.3764810}
}

@inproceedings{jenga,
  title={Jenga: Effective Memory Management for Serving {LLM} with Heterogeneity},
  author={Chen Zhang and Kuntai Du and Shu Liu and Woosuk Kwon and Xiangxi Mo and Yufeng Wang and Xiaoxuan Liu and Kaichao You and Zhuohan Li and Mingsheng Long and Jidong Zhai and Joseph E. Gonzalez and Ion Stoica},
  booktitle={Proceedings of the ACM SIGOPS 31st Symposium on Operating Systems Principles},
  pages={446--461},
  year={2025},
  publisher={Association for Computing Machinery},
  doi={10.1145/3731569.3764823},
  url={https://doi.org/10.1145/3731569.3764823}
}

@inproceedings{shadowkv,
  title={{ShadowKV}: {KV} Cache in Shadows for High-Throughput Long-Context {LLM} Inference},
  author={Hanshi Sun and Li-Wen Chang and Wenlei Bao and Size Zheng and Ningxin Zheng and Xin Liu and Harry Dong and Yuejie Chi and Beidi Chen},
  booktitle={Proceedings of the 42nd International Conference on Machine Learning},
  series={Proceedings of Machine Learning Research},
  volume={267},
  pages={57355--57373},
  year={2025},
  publisher={PMLR},
  url={https://proceedings.mlr.press/v267/sun25b.html}
}

@article{tokendance,
  title={{TokenDance}: Scaling Multi-Agent {LLM} Serving via Collective {KV} Cache Sharing},
  author={Zhuohang Bian and Feiyang Wu and Chengrui Zhang and Hangcheng Dong and Yun Liang and Youwei Zhuo},
  journal={arXiv preprint arXiv:2604.03143},
  year={2026},
  url={https://arxiv.org/abs/2604.03143}
}

@inproceedings{kvflow,
  title={{KVFlow}: Efficient Prefix Caching for Accelerating {LLM}-Based Multi-Agent Workflows},
  author={Zaifeng Pan and Ajjkumar Patel and Yipeng Shen and Zhengding Hu and Yue Guan and Wan-Lu Li and Lianhui Qin and Yida Wang and Yufei Ding},
  booktitle={Advances in Neural Information Processing Systems},
  volume={38},
  year={2025},
  doi={10.52202/085713-4208},
  url={https://proceedings.neurips.cc/paper_files/paper/2025/hash/b7971d31a7d5eb0f1eed2f8f6f368195-Abstract-Conference.html}
}

\end{document}